\documentclass[aps,prd,twocolumn,superscriptaddress,nofootinbib,10pt]{revtex4-2}

\usepackage[utf8]{inputenc}
\usepackage{graphicx}
\usepackage[separate-uncertainty=true,exponent-product=\cdot]{siunitx}
\usepackage{amsmath}
\usepackage{amssymb}
\usepackage{amsfonts}
\usepackage{bm}
\usepackage{xcolor}
\usepackage{mathtools}
\usepackage{standalone}
\usepackage{tikz}

\usepackage[colorlinks=true, allcolors=darkblue]{hyperref} 
\definecolor{darkblue}{rgb}{0,0.2,0.6}

\usetikzlibrary{arrows.meta, positioning, calc}

\newcommand{\medianeff}{1\%}
\newcommand{\samplespersec}{5000}
\newcommand{\issamplespersec}{800}
\newcommand{\mediantimetothousandneff}{\SI{2}{min}}
\newcommand{\cpudays}{100}
\newcommand{\secpersimulation}{0.9}
\newcommand{\labradorfaster}{$1000 \times$}
\newcommand{\lowmassrange}{(1, 15)}
\newcommand{\highmassrange}{(12, 50)}
\newcommand{\qrange}{(0.1, 1)}
\newcommand{\effrawweights}{0.011}
\newcommand{\effcounterweights}{0.42}
\newcommand{\effratio}{40}

\newcommand{\rmd}{\mathrm{d}}
\newcommand{\rmi}{\mathrm{i}}
\newcommand{\rme}{\mathrm{e}}

\newcommand{\rml}{\mathrm{L}}
\newcommand{\los}{\bm{\hat n}}

\DeclareMathOperator*{\mean}{mean}
\DeclareMathOperator*{\std}{std}
\DeclareMathOperator*{\var}{var}

\newcommand{\tapir}{\affiliation{TAPIR, California Institute of Technology, Pasadena, CA 91125, USA}}
\newcommand{\LIGOlab}{\affiliation{LIGO Laboratory, California Institute of Technology, Pasadena, California 91125, USA}}

\definecolor{niceblue}{RGB}{44,110,160}
\definecolor{nicegreen}{RGB}{60,160,40}
\definecolor{nicered}{RGB}{180,70,60}

\colorlet{bluefill}{niceblue!15}
\colorlet{bluedraw}{niceblue!70!black}
\colorlet{greenfill}{nicegreen!15}
\colorlet{greendraw}{nicegreen!70!black}
\colorlet{redfill}{nicered!15}
\colorlet{reddraw}{nicered!70!black}

\makeatletter
\newcommand{\subalign}[1]{%
  \vcenter{%
    \Let@ \restore@math@cr \default@tag
    \baselineskip\fontdimen10 \scriptfont\tw@
    \advance\baselineskip\fontdimen12 \scriptfont\tw@
    \lineskip\thr@@\fontdimen8 \scriptfont\thr@@
    \lineskiplimit\lineskip
    \ialign{\hfil$\m@th\scriptstyle##$&$\m@th\scriptstyle{}##$\hfil\crcr
      #1\crcr
    }%
  }%
}
\makeatother

\graphicspath{ {figures/} }

\begin{document}

\title{%\texttt{labrador}:
A domain-optimized machine-learning tool for gravitational wave inference}

\author{Javier Roulet}
\email{jroulet@uchicago.edu}
\tapir
\affiliation{Kavli Institute for Cosmological Physics, The University of Chicago, 5640 South Ellis Avenue, Chicago, Illinois 60637, USA}
\affiliation{School of Natural Sciences, Institute for Advanced Study, 1 Einstein Drive, Princeton, NJ 08540, USA}

\author{Marco Crisostomi}
\tapir
\affiliation{Dipartimento di Fisica, Universit\`a di Pisa, Largo B. Pontecorvo 3, 56127 Pisa, Italy}

\author{Lucy M.\ Thomas}
\tapir
\LIGOlab

\author{Katerina Chatziioannou}
\tapir
\LIGOlab

\begin{abstract}
Fast and reliable inference of gravitational-wave source parameters is crucial for analyzing large catalogs that are reaching the size of hundreds of detections, and for identifying short-lived electromagnetic counterparts. 
Neural posterior estimation has emerged as a powerful inference method, where the model is trained on simulated gravitational-wave data at considerable computational cost, but thereafter enables extremely fast and inexpensive inference at test time.
Here, we extend this approach by incorporating domain-specific physical insights and methods in the model architecture.
These include compressing the data by heterodyning against a reference waveform chosen via approximate likelihood maximization, removing parameter degeneracies through tailored coordinate systems, and eliminating known multimodalities by folding the parameter space.
As a result, the network is approximately equivariant to changes in the source parameters, and achieves a reduced training cost and improved model interpretability.
Our implementation, called \texttt{labrador},\footnote{\label{ftn:github}%
    Likelihood-free
    Amortized
    Bayesian
    Retrieval
    Applying
    Domain-%
    Optimized
    Representations,
    \url{https://github.com/jroulet/labrador}
} can be trained end-to-end on a 1-day timescale on $\sim 10^2$ CPU cores and a V100 GPU, achieving a median importance-sampling efficiency of {\medianeff} on quadrupolar, aligned-spin signals in a broad mass range (chirp mass $\mathcal M \in 1\text{--}50\,\mathrm{M}_\odot$, mass ratio $q > 0.1$).
\texttt{labrador} is the first neural inference code to achieve extensive coverage of long-duration signals with secondary masses $m_2 < \SI{10}{M_\odot}$, rendered possible by its equivariance property.
Among our novel contributions is a numerically stable procedure that enables neural posterior estimation when the simulation and inference priors differ.
\end{abstract}

\maketitle

\section{Introduction}
\label{sec:introduction}

LIGO--Virgo--KAGRA have observed signals from mergers of binary neutron stars \cite{Abbott2017, Abbott2020, Niu2025}, neutron-star--black-hole binaries \cite{Abbott2021, Abac2024}, and hundreds of binary black holes \cite{Abac2026, Nitz2023, Wadekar2025}.
The number of detections is forecasted to continue increasing about tenfold every five years \cite{Broekgaarden2024} thanks to instrument upgrades  \cite{Aasi2015, Acernese2014, Akutsu2019, Reitze2019, Hild2011, Punturo2010, Capote2025}.
Efficient and accurate parameter inference is critical to enable the full scientific potential of these data, including prompt followup of transient electromagnetic counterparts and controlling systematic errors as catalogs grow, as well as for making forecasts and simulating mock datasets.

While traditional inference techniques like Markov-chain Monte Carlo \cite{Metropolis1953, Hastings1970} and nested sampling \cite{Skilling2006} are mature in gravitational-wave analysis \cite{Veitch2015, Lange2018, Ashton2019, Ashton2021, Biwer2019, Breschi2021, Cornish2021quickcbc, Roulet2022, Fairhurst2023, Tiwari2023, Wong2023}, they remain slow ($> \SI{1}{\hour}$ per event for complex waveform models).
Emerging machine-learning methods for amortized inference, such as conditional neural posterior estimation, produce posteriors in seconds by training on synthetic data beforehand~\cite{Chua2020, Green2020, Gabbard2021, Green2021, Dax2021, Dax2022, Dax2023, Wildberger2023b, Chatterjee2024, Dax2025, Kofler2025, Raymond2025, Spadaro2026, AlShammari2026}.
Beyond speed at inference time, these algorithms offer other unique advantages.
Representations of the posterior distribution such as normalizing flows \cite{Rezende2015}, diffusion models \cite{Ho2020} or flow matching \cite{Lipman2023} are more powerful than samples---they can generate an arbitrary number of samples efficiently on demand, but also evaluate their (correctly normalized) density.
Samples produced in this way have a firm guarantee of obeying the model distribution (while stochastic sampling algorithms occasionally fail to converge), which enables extremely powerful diagnostics of convergence by importance sampling \cite{Dax2023}.
Looking forward, these representations can also be more lightweight than storing large numbers of samples explicitly, even acting as sufficient statistics for downstream analyses \cite{Leyde2024}.
Samples need not be generated independently, but could be chosen on a low-discrepancy sequence to reduce the variance of importance-sampling estimates \cite{Andral2024} such as used in hierarchical Bayesian inference \cite{Mandel2019, Thrane2019, Roulet2020}.
Finally, implicit-likelihood methods offer a path towards optimal inference in the absence of a computable likelihood function, such as with non-Gaussian or non-stationary noise \cite{Legin2025,Raymond2025,Negri2026,Emma2026}.
On the other hand, the main challenges that these approaches currently face involve training complexity, robustness and accuracy, which demand long training times and corrections in postprocessing.

In this paper, we set out to improve the training efficiency of simulation-based inference for gravitational-wave sources by adapting domain-specialized techniques and insights.
A guiding principle in our design is to build representations of the strain data and parameter space that are approximately invariant under changes in the source parameters, which significantly reduces and organizes the variability in the training set.
As a result, the number of trainable parameters can be made significantly smaller, and the model is applicable throughout parameter space regardless of the signal duration---a limiting factor for many state-of-the-art models.
Throughout most of this work we will adopt the simplifying assumptions that the noise is Gaussian and stationary with a fixed and known power spectral density (PSD), and that the signals consist of quadrupolar radiation from aligned-spin (non-precessing) compact binary mergers.
Having demonstrated the worth of domain-specific optimizations, we plan to increase model complexity in the future, including higher-order modes, precessing spins, eccentricity, varying PSD, etc.

We structure the paper as follows:
In \S\ref{sec:data_representation} we introduce a data representation that is approximately invariant to a broad class of signal variability, thereby making the downstream model naturally equivariant.
In \S\ref{sec:parameter_representation} we introduce a series of reparametrizations that simplify the characteristic structure of gravitational-wave posteriors.
In \S\ref{sec:training_set} we describe the construction of our low-discrepancy training set, and develop a novel technique for amortized inference with mismatched simulation and training priors.
In \S\ref{sec:convergence} we report our results in terms of training cost and model performance.
We conclude, and compare our results to other current models, in \S\ref{sec:conclusions}.

%%%%%%%%%%%%%%%%%%%%%%%%%%%%%%%%%%%%%%%%%
\section{Data representation}
\label{sec:data_representation}
%%%%%%%%%%%%%%%%%%%%%%%%%%%%%%%%%%%%%%%%%

A major challenge in gravitational-wave inference is the large size of the signal space. 
The intrinsic parameters (masses and spins) account for $> 10^5$ independent waveform shapes (as reflected in the size of template banks used for detection) even in the simplified case of aligned-spin, quadrupole-only, circular-orbit signals \cite{Roulet2019, Roy2019}, and one-to-two orders of magnitude more for signals with generic spin \cite{Schmidt2024b}, higher modes \cite{Harry2018, Chandra2022, Schmidt2024} or eccentricity \cite{Phukon2025}.
On top of this, extrinsic parameters (location, orientation, time) greatly add to the variability as they affect the very-well measured arrival time, amplitude and phase at each detector.
Indeed, the reason why a statistically significant detection of a gravitational-wave requires a fairly large signal-to-noise ratio $\gtrsim 8$ is that so many independent signal templates are tried that Gaussian noise is bound to produce $8\sigma$ fluctuations \cite{Babak2013}.

This said, gravitational-wave chirps exhibit a very characteristic morphology, and thus lie in a comparatively small manifold within the space of all possible signals.
In this section we develop a representation of gravitational-wave strain data that is tailored to this structure, enabling efficient compression, approximate invariance, and improved interpretability.

%%%%%%%%%%%%%%%%%%%%%%%%%%%%%%%%%%%%%%%%%%%%%
\subsection{Source-parameter variations viewed as an equivariance group}

To a good---but not exact---approximation, the dependence of the gravitational-wave strain on the parameters of the source has a simple analytical form.
We aim to incorporate this knowledge into the model architecture, so that the neural network does not need to rediscover this structure from the training simulations, but instead focuses on learning departures from the analytical model.

A way of formalizing the connection between source parameters and strain data is to treat parameter variations as elements of a transformation group.
The group actions on these two spaces are different: for example, time-translating a waveform amounts to permuting values in the timeseries, while time-translating the source parameters corresponds to incrementing a time variable.

One can incorporate this structure into the model by guaranteeing that the outputs transform consistently under transformations of the inputs, in which case the model is said to be equivariant under the transformation group.
One way of enforcing these symmetries is through a careful design of the neural network architecture, a classic example being convolutional neural networks for translational equivariance \cite{Lecun1998,Cohen2016}.
A simpler approach, which we will adopt in this work, is transforming the inputs to make them invariant. 
This procedure is called standardizing the pose of the data and makes all downstream stages of the model invariant regardless of their architecture.

This approach was introduced to gravitational-wave inference by \citet{Dax2022}, who proposed using group transformations (e.g., time shifts) to map waveforms to a standardized reference pose. 
This allows to split the problem of representing the data into two tasks: finding the ``correct'' translation that standardizes the pose of the data (arrival time near zero), and describing the standardized data, which exhibits less variability and is therefore easier to compress and analyze.
In their implementation, these steps are applied iteratively to reach convergence.
The transformation group may correspond either to an exact or approximate equivariance.

Building on this idea, in this section we extend the transformation group to a much broader class of equivariances that cover all ``major" parameters of the gravitational-wave signal.
These include the time, amplitude, and phase of arrival at each detector, as well as the leading-order terms in the post-Newtonian expansion that govern the rate of orbital decay.
As a result, the majority of the intrinsic variability of the signal is eliminated from the data representation---or, rather, distilled into a low-dimensional set of interpretable parameters that define the standardizing transformation.
Importantly, because the pose-standardized data preserve any residual information not accounted for by this transformation, the transformation itself need not constitute an exact model of the data; in such cases, the associated invariance is only approximate.
This allows us to work with approximate, analytical waveform models for the purpose of standardizing the data.

A frequency-domain gravitational-wave inspiral can be analytically approximated as
\begin{align}
    \label{eq:phenom_waveform}
    \hat h_k(f; \bm p) &=
        A_k f^{-7/6} \rme^{\rmi \Phi_k(f; \bm p)}\,, \\
    \Phi_k(f; \bm p) &= 
        \varphi_k
        - 2 \pi f t_k
        + \sum_n a_n f^{\gamma_n}
    \\ &= 
        \overline \Phi(f) +
        \sum_\alpha c_\alpha(\bm p) e_{k \alpha}(f)\,, \label{eq:calpha}\\
    \bm p &= \big(
        \{
            A_k, \varphi_k, t_k
        \}_{k=1}^{N_\mathrm{det}},
        \{a_n\}_{n \in \textrm{PN}} \big)\,.
\end{align}
We place a caret on $\hat h$ to acknowledge that this model is not exact, and is only supposed to describe waveforms approximately.
Rather than using the physical parameters of the source, we define phenomenological parameters $\bm p$ that include the amplitude $A_k$, phase $\varphi_k$ and time $t_k$ of arrival at each detector $k$, which depend on the extrinsic parameters of the source, as well as a small set of coefficients of the post-Newtonian (PN) expansion  $\{a_n\}$, which describe the underlying shape of the waveform and are related to the intrinsic parameters \cite{Blanchet2002}.
The amplitude, phase and time observed at each detector are different, since they depend on the orientation and location of each interferometer relative to the source, while the intrinsic shape of the waveform is the same across detectors and so the $a_n$ are shared.
To further simplify the description, in Eq.~\eqref{eq:calpha} we re-express the waveform phase in terms of orthonormal (with respect to the mismatch metric) functions $e_{k\alpha}(f)$ and coefficients $c_\alpha$, which are linear combinations of the elements of $\bm p$ (a similar decomposition is used in the construction of ``geometric'' template banks \cite{Brown2013,Roulet2019}).
This orthonormalization will prove useful later, both for optimizing the reference waveform efficiently and for decorrelating the features passed to the network.
Appendix \ref{app:waveform} describes how these phenomenological parameters relate to the physical parameters and how the basis elements $e_{\alpha}$ are chosen, and defines all quantities.

Equation~\eqref{eq:phenom_waveform} shows that valid waveforms are related by simple transformations.
Namely, under a change in the parameters the waveform transforms according to:
\begin{subequations}
    \label{eq:transform}
    \begin{align}
        \delta A_k \quad&\Rightarrow\quad
            \hat h_k \mapsto (1 + \delta A_k / A_k) \hat h_k\,,
            \label{eq:amp_transform}\\
        \delta c_\alpha \quad&\Rightarrow\quad
            \hat h_k \mapsto \rme^{\rmi \, \delta c_\alpha e_{k\alpha}(f)} \hat h_k\,.
            \label{eq:coef_transform}
    \end{align}
\end{subequations}
Since Eq.~\eqref{eq:phenom_waveform} is only an approximate phenomenological description, most of these equivariances are not exact for real signals.
Another subtlety is that, while we introduced these as symmetries of the waveform, in truth it is the noisy data whose pose we are interested in standardizing.
The transformations of Eq.~\eqref{eq:coef_transform} are symmetries of Gaussian noise, in the sense that they map one realization of noise into another that is equally probable, and furthermore preserve the distribution.
This is because Gaussian noise satisfies $p(n(f)) = p(\rme^{\rmi \, \delta\Phi(f)}n(f))$ for any $\delta\Phi(f)$.
Instead, rescaling the amplitude per Eq.~\eqref{eq:amp_transform} is not a symmetry of the noise, and so, for this reason too, this transformation is not an exact equivariance of the data.

In summary, the data are composed of noise, plus a signal that can be approximated by Eq.~\eqref{eq:phenom_waveform} with best-fit parameters $\bm p_*$.
We will interpret $\bm p_*$ as a pose-standardizing transformation, and represent the data in terms of $\bm p_*$ and the pose-standardized data.
The parameters $\bm p_*$ capture most of the signal variability, while the standardized data retain any information not encoded in the reference waveform $\hat h_k(\bm p_*)$ and are otherwise invariant to the signal parameters.
Next, we describe this procedure in detail.

%%%%%%%%%%%%%%%%%%%%%%%%%%%%%%%%%%%%%%%%%%%%%%%%
\subsection{Optimizing the reference parameters}

A core component of the data representation are the best-fit parameters $\bm p_*$ that define the standardizing transform.
We obtain these by maximizing the likelihood.
Since this step depends on the observed data, it is not amortized and needs to be performed very efficiently for each event after detection.
Moreover, since the optimization influences the data representation, it also needs to be performed when creating the training set, and in fact dominates the computational cost of generating training data in our approach.
(That this is the case has one drawback: it prevents  us from varying the extrinsic parameters and noise realizations on the fly during training---an approach used by other codes to diversify the training set and reduce overfitting \cite{Dax2021}.)

The space of phenomenological parameters $\bm p$ is $\sim 10$-dimensional ($3N_\mathrm{det} + N_\textrm{PN}$), however, the amplitude and phase at each detector can be fitted analytically \cite[e.g.][\S IV]{Allen2012}, which accounts for $2N_\textrm{det}$ of those parameters.
We optimize the remaining ones with a global maximization algorithm (differential evolution \cite{Storn1997, Virtanen2020}), initialized around a guess that the user gives in the form of a reference phase $\Phi^\textrm{guess}_k(f)$ (e.g., that of the template that triggered the detection).
For simplicity, in the training set we instead initialize the optimizer with the phase profile of the injected signal.
Importantly, the maximization step erases this information, as desired because in real data it is inaccessible.
The optimization is simplified by our orthonormal coordinates $\bm c$, in which the likelihood function resembles an isotropic multivariate Gaussian.
With this implementation, finding the reference waveform takes $\approx \secpersimulation$\,s per event on one CPU core.

%%%%%%%%%%%%%%%%%%%%%%%%%%%%%%%%%%%%%%%%%%%%%%%%
\subsection{Heterodyning as a standardizing transform}
\label{sec:heterodyning}

\begin{figure}
    \centering
    \includegraphics[width=0.9\linewidth]{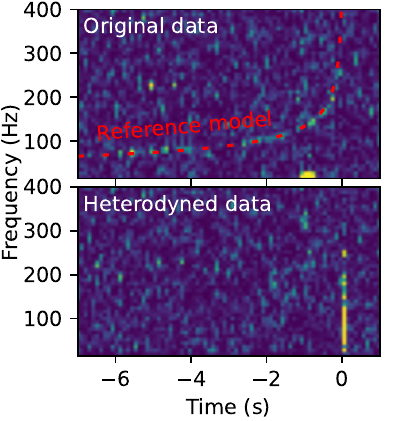}
    \caption{Whitened strain data for GW170817, before and after standardizing its pose by heterodyning against the reference waveform model of Eq.~\eqref{eq:phenom_waveform}.
    The heterodyned data are approximately invariant to the parameters of the signal.
    Furthermore, they can be compressed very efficiently:
    the signal power gets concentrated from minutes to milliseconds, allowing to retain a small time interval or, equivalently, reduce the frequency resolution. }
    \label{fig:heterodyned}
\end{figure}

Having found the best-fit $\bm p_*$, we standardize the pose of the data by applying the inverse of the transform  \eqref{eq:transform}:
\begin{equation}
    \label{eq:heterodyne}
    d_k'(f) = \frac{d_k(f)}{\hat h_k(f; \bm p_*)}\,,
\end{equation}
using Eqs.~\eqref{eq:phenom_waveform} and \eqref{eq:calpha}.
To the extent that the reference waveform describes well the data, the pose-standardized data $d'$ is unity, clearly invariant to the signal parameters.
This is illustrated in Fig.~\ref{fig:heterodyned}, which shows a time-frequency representation of the whitened strain data for GW170817 \cite{Abbott2017} before and after standardizing its pose.
Equation~\eqref{eq:heterodyne} is essentially the same as heterodyning, or ``dechirping'' the data with a filter $1/\hat h_k^*(f; \bm p_*)$ (a standard technique in transient gravitational-wave data analysis \cite{Cornish2013, Zackay2018, Cornish2021, Nitz2026}), so we will refer to $d'$ as the heterodyned data.
Given the reference waveform and the heterodyned data, it is possible to reconstruct the original data---there is no loss of information even though the reference model is approximate.

The heterodyned data are not only approximately invariant to the signal parameters, but also easily compressed by reducing the frequency resolution (equivalently, cropping a small time segment in Fig.~\ref{fig:heterodyned}).
We do this in a principled way by means of the relative-binning algorithm \cite{Zackay2018}.
The crux is that all the information in the data about the source parameters must be encoded in the likelihood function
\begin{equation}
    \label{eq:likelihood}
    \ln \mathcal L(d \mid \theta) \sim
        \langle d \mid h \rangle
        - \frac 12 \langle h \mid h \rangle\,,
\end{equation}
(where $\langle \cdot \mid \cdot \rangle$ is the inverse-variance-weighted inner product, defined as in Eq.~\eqref{eq:relative_binning} below)---and, specifically, in the $\langle d \mid h \rangle$ term, since $\langle h \mid h \rangle$ is independent of the data (other than through the noise power spectrum, which for now we approximate as fixed).
Thus, it suffices to provide the network with enough information to reconstruct $\langle d \mid h \rangle$.
Relative-binning (as implemented in \citet{Roulet2024}) evaluates this term from precomputed weights $w_{kb}$ as
\begin{equation}
    \label{eq:relative_binning}
    \begin{split}
    \langle d \mid h \rangle 
    &= \sum_{k\in \text{det}}
        4 \, \Re\int_0^\infty \rmd f
        \frac{d_k(f) h_k^*(f)}{S_k(f)} \\
    &\approx \Re\sum_{k,b}
        \frac{h_k^*(f_b)}{\hat h_k^*(f_b; \bm p_*)}
        w_{kb}\,;\\
    w_{kb} &= 4 \int_0^\infty \rmd f \frac{d_k(f) \hat h_k^*(f; \bm p_*)}{S_k(f)}s_b(f)\,,
    \end{split}
\end{equation}
where $\{f_b\}$ is a coarse set of $\sim 100$ frequencies (a suitable choice based on the post-Newtonian expansion is given in Ref. \cite{Zackay2018}), and $\{s_b(f)\}$ are a set of splines with the defining property that $s_b(f_{b'}) = \delta_{bb'}$.
It follows from Eq.~\eqref{eq:relative_binning} that the weights $w_{kb}$ contain all the relevant information in the data. 
Furthermore, using Eqs.~\eqref{eq:phenom_waveform} and \eqref{eq:heterodyne}, these weights can be expressed in terms of the heterodyned data as $w_{kb} = A_{k*}^2 d'_{kb}$, where $A_{k*} \in \bm p_*$ is the best-fit amplitude in the $k$th detector and
\begin{equation}
    \label{eq:heterodyned_data}
    d_{kb}' =
        4\int_0^\infty \rmd f\,
        d'_k(f)
        \frac{f^{-7/3}}{S_k(f)}
        s_b(f)
\end{equation}
is a low-frequency-resolution projection of the heterodyned data.
$d'_{kb}$ have the desirable properties of being approximately invariant to the source parameters, highly compressed, and containing, together with $\bm p_*$, all the information available about the source parameters.
Moreover, they act as a way to standardize the dimensionality of the data across events of different duration (i.e., frequency resolution).
During training, we use segments of \SI{32}{\second}, long enough to resolve narrow-band features in the noise PSD, and use the frequency range $15$--\SI{1024}{Hz}.
We do not apply any windowing, so that the Gaussian noise is periodic and any signal longer than the segment simply wraps around (the heterodyning step later undoes this), which is effectively equivalent to having arbitrarily long training data segments.
At inference time, we use the same interpolation nodes $\{f_b\}$ as in training, and compute the basis splines $\{s_b(f)\}$ on the arbitrary frequency grid of the data.

\citet{Dax2025,Hu2025} have also implemented heterodyning as a form of data compression in the context of simulation-based inference.
They choose their reference from a one-dimensional grid of chirp masses, by trying all possible references, generating a small set of posterior samples with each, and then adopting the reference that yields the sample with the highest likelihood.
This strategy may be challenging to scale to higher-dimensional families of reference waveforms, as the associated space may be too large to tile efficiently.
In contrast, our procedure (global optimization, aided by analytical maximization over detectors' amplitude and phase, and by coordinate orthonormalization) scales satisfactorily at least up to $\sim 10$ dimensions.
An additional advantage is that our reference parameters are continuous-valued, and thus the pose of the data (defined through the reference waveform) gets standardized more accurately.
All in all, the maximization step allows us to circumvent the chirp-mass optimization over a grid, as well as the group-equivariant neural posterior estimation architecture of \citet{Dax2022} with respect to the time of arrival (that method has the drawback of losing access to the posterior density), while making the representation itself equivariant to a much larger class of approximate symmetries.
We have also pointed out a conceptual connection with group equivariance: heterodyning against a reference waveform corresponds to standardizing the pose of the data with respect to all the parameters of the reference waveform.

%%%%%%%%%%%%%%%%%%%%%%%%%%%%%%%%%%%%%%%%%%%%%%%
\subsection{Compression via singular value decomposition}
\label{sec:compression}

Even at coarse frequency resolution, the heterodyned data remain several-hundred dimensional. 
In this subsection we seek an even lower dimensional representation that still preserves the signal information.
By the arguments in \S\ref{sec:heterodyning}, the heterodyned data are approximately invariant to the signal parameters, and the deviations from invariance encode features missing in the phenomenological model of Eq.~\eqref{eq:phenom_waveform}.
These include detector noise and refinements of the waveform model.
Our goal is to retain the latter at the expense of the former while further compressing the heterodyned data, which we achieve using a singular value decomposition (SVD).
While an SVD has also been used by \citet{Dax2021, Dax2025}, we detail our procedure below since it has a different whitening step, and it introduces a principled criterion for selecting the dimensionality.

The heterodyned data in Eq.~\eqref{eq:heterodyned_data} form a tensor  
\[
d' \in \mathbb{C}^{N_\mathrm{sim} \times N_\mathrm{det} \times N_\mathrm{freq}}\,,
\]  
with complex values over a few hundred frequencies for each detector and simulation.
We reshape $d'$ by concatenating the real and imaginary parts from all detectors:
\[
(d'_{ij}) \in \mathbb{R}^{N_\mathrm{sim} \times (2 N_\mathrm{det} N_\mathrm{freq})}\,,
\]  
with $i$ indexing the simulation and $j$ the combined real/imaginary–detector–frequency axis.
In the remainder of this section, we assume all data and signals are heterodyned and drop the prime notation.
Likewise, we define the heterodyned noiseless signals (heterodyned with the same reference waveforms found by the maximizer in the noisy data), which we call $h_{ij}$, and the residual $n_{ij} = d_{ij} - h_{ij}$.
Neither $h_{ij}$ nor $n_{ij}$ are observable in real data, since they depend on the true signal; we will only use them to identify, via SVD, the linear combinations of the heterodyned data that are most informative about the signals.

Since SVD defines principal components according to an Euclidean metric, we precondition the heterodyned data by centering and whitening it:
\begin{align}
    d^{\rm w}_{ij} &= \frac{d_{ij} - \overline h_j}{\sigma_j}\,, \\
    h^{\rm w}_{ij} &= \frac{h_{ij} - \overline h_j}{\sigma_j}\,, \\
    n^{\rm w}_{ij} &= \frac{n_{ij}}{\sigma_j}\,,
\end{align}
where
\begin{align}
    \overline h_j &= \mean_i h_{ij}\,, \\
    \sigma_j &= \std_i n_{ij}\,.
\end{align}
Because we work with heterodyned signals (Eq.~\eqref{eq:heterodyned_data}), the whitening procedure has two differences with that in \citet{Dax2021}: first, $\overline h_{j} \not\approx 0$, and thus centering is necessary;
second, the dispersion $\sigma_j$ is not the amplitude spectral density of the detector noise.
There are some subtleties that we ignore, for example, the noise and signals get correlated by performing the maximization, and the noise has a small but non-zero mean because the phenomenological model Eq.~\eqref{eq:phenom_waveform} cannot recover all the signal content.
We perform an SVD of the whitened noiseless signals:
\begin{equation}
    h^{\rm w}_{ij} = \sum_\alpha U_{i\alpha} D_\alpha V^\dagger_{\alpha j}\,,
\end{equation}
where $U$ and $V$ are orthogonal matrices.
The rows of $V^\dagger$ define an orthonormal basis in the whitened data space, aligned with the directions where the noiseless signals have the largest variance.
We store the $V$ matrix, since it will be necessary to compress any future data after the network has been trained.
The SVD-compressed data have the form
\begin{equation}\label{eq:svdcoef}
    d^\mathrm{SVD}_{\alpha}
    = \sum_j
        d^{\rm w}_{j} V_{j \alpha}\,, \quad 1 \leq \alpha \leq \alpha_\mathrm{max}\,;
\end{equation}
the compression consists of using only the first few $\alpha$ (with the largest eigenvalues $D_\alpha$) and discarding the rest.
This entails a tradeoff between preserving information and reducing dimensionality.

\begin{figure}
    \centering
    \includegraphics[width=\linewidth]{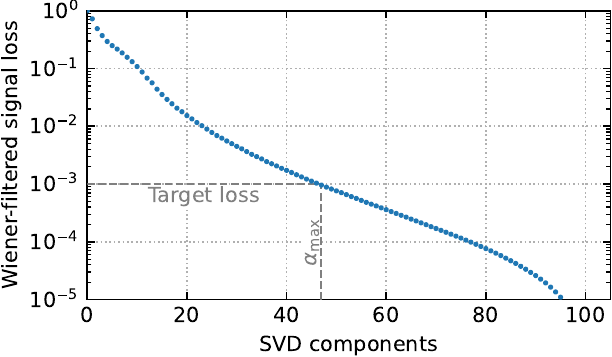}
    \caption{Efficient compression of the heterodyned data via SVD. In this example, the heterodyned data from two detectors can be compressed into under $50$ real components while capturing $99.9\%$ of the power of the Wiener-filtered signal.
    The plot shows $\sum_{\alpha>\alpha_\mathrm{max}}W^2_\alpha S_\alpha^{(h)} / \sum_{\alpha}W^2_\alpha S_\alpha^{(h)}$ versus $\alpha_\mathrm{max}$.
    }
    \label{fig:compression_loss}
\end{figure}

We determine the number of SVD coefficients to retain, $\alpha_\textrm{max}$, by requiring that the Wiener-filtered signal can be reconstructed to a specified accuracy, illustrated in Fig.~\ref{fig:compression_loss}.
The Wiener filter is the optimal linear estimator of a signal of interest in the presence of noise \cite{Wiener1949}.
In our case, the signal is $h^\mathrm{SVD}_{\alpha}$ and the observable is $d^\mathrm{SVD}_{\alpha}$.
We construct the Wiener filter $W_\alpha$ from the spectrum of the signals and noise as
\begin{align}
    W_\alpha
    &= \frac{S^{(h)}_\alpha}
            {S^{(h)}_\alpha + S^{(n)}_\alpha}\,; \\
    S^{(h)}_\alpha &= \var_i h^\mathrm{SVD}_{i\alpha}\,, \\
    S^{(n)}_\alpha &= \var_i n^\mathrm{SVD}_{i\alpha}\,.
\end{align}
where $h^\mathrm{SVD}_{i\alpha}$ and $n^\mathrm{SVD}_{i\alpha}$ are found as in Eq.~\eqref{eq:svdcoef} but using the whitened signals and noise, respectively.
The filtered data $W_\alpha d^\mathrm{SVD}_{\alpha}$ have a spectrum $W^2_\alpha (S^{(h)}_\alpha + S^{(n)}_\alpha)$.
We truncate the representation at an $\alpha_{\rm max}$ such that 99.9\% of the signal power $W^2_\alpha S^{(h)}_\alpha$ is recovered; we find that for two detectors this is typically achieved with $\approx 50$ components.
For comparison, a two-minute long binary neutron star signal sampled at \SI{2}{\kilo\hertz} in two detectors would require $2^{19} \approx \num{5e5}$ floating point numbers to represent as a time series, a factor \num{e4} larger.
The reduced size of our representation facilitates the use of smaller neural networks and enables loading the entire dataset ($<\SI{3}{GB}$ for \num{e7} simulations) even on modest GPU resources.

\begin{figure*}
    \centering
    \begin{tikzpicture}[
    bluebox/.style={
    rectangle, rounded corners=4pt,
    draw=bluedraw, fill=bluefill,
    minimum width=2.0cm, minimum height=0.65cm,
    align=center, font=\small
    },
    redbox/.style={
    rectangle, rounded corners=4pt,
    draw=reddraw, fill=redfill,
    minimum height=0.65cm,
    align=center, font=\small
    },
    greenbox/.style={
    rectangle, rounded corners=4pt,
    draw=greendraw, fill=greenfill,
    minimum height=0.65cm,
    align=center, font=\small
    },
    arr/.style={-{Stealth[length=6pt, width=5pt]}, thick, color=black!70},
    darr/.style={-{Stealth[length=6pt, width=5pt]}, semithick, dashed, color=black!50},
]
% Bottom row
\node[bluebox] (dk) at (0,0)    {$d_k(f)$};
\node[bluebox] (dkdash)  at (9.0,0)  {$d'_k(f)$};
\node[bluebox] (dkbdash)  at (12.0,0.0)  {$d'_{kb}$};
\node[bluebox] (dijw)  at (15.0,0.0)  {$d^{\text{w}}_{j}$};
\node[bluebox] (dsvd) at (18.0,0.0) {$d_{\alpha}^{\text{SVD}}$};
\node[greenbox] (d) at (20.0,0.0) {$d$};

% Upper row
\node[bluebox] (pstar)  at (2.0,2.0)  {$\bm{p}_{*}$};
\node[bluebox] (hcaret)  at (5.0,1.0)  {$\hat{h}_{k}(f;\bm{p}_{*})$};

% --- Arrows bottom row ---
%\draw[arr] (dk) |- (pstar);
\draw[arr] (dk) |- node[pos=0.5, left, font=\small, align=center] {likelihood\\maximization} (pstar);
%\draw[arr] (pstar.south) |- (hcaret);
\draw[arr] (pstar.south) |- node[pos=0.7, below, font=\small] {Eqs.~(1)--(2)} (hcaret);
\draw[arr] (hcaret.south) -| ($(dk)!0.555!(dkdash)$);
%\draw[arr] (hcaret) -| (dkdash);
%\draw[arr] (dk) -- (dkdash);
\draw[arr] (dk) -- node[pos=0.8, below, font=\small, align=center] {heterodyning\\Eq.~(6)} (dkdash);
\draw[arr] (dkdash) -- node[pos=0.5, above, font=\small,yshift=6pt,align=center] {coarsening\\Eq.~(9)} (dkbdash);
\draw[arr] (dkbdash) -- node[pos=0.5, above, font=\small,yshift=6pt,align=center] {centering, whitening\\Eq.~(10)} (dijw);
\draw[arr] (dijw) -- node[pos=0.5, above, font=\small,yshift=6pt,align=center] {SVD\\Eq.~(16)} (dsvd);
\draw[arr] (dsvd) -- (d);
\draw[arr] (pstar) -| (d);
%\draw[arr] (rescaler) -- (thetaF);
%\draw[arr] (thetaF) -- (ell);
%\draw[arr] (ell) -- (thetaT);
%\draw[arr] (theta) -- (thetaT);

% --- Vertical arrows ---
%\draw[arr] (d) -| (sbi);

\end{tikzpicture}
    \caption{Construction of our data representation.
    A phenomenological waveform $\hat h_k(d; \bm p_*)$ is fitted to the observed strain $d_k(f)$, and used to heterodyne them ($d'_k(f)$).
    The heterodyned data are recast at reduced frequency resolution ($d'_{kb}$), whitened with respect to their empirical mean and variance ($d_j^\mathrm{w}$) and projected onto the SVD basis ($d_\alpha^\mathrm{SVD}$).
    The data are represented through the leading SVD components together with the parameters of the reference waveform.
    }
    \label{fig:data_representation}
\end{figure*}
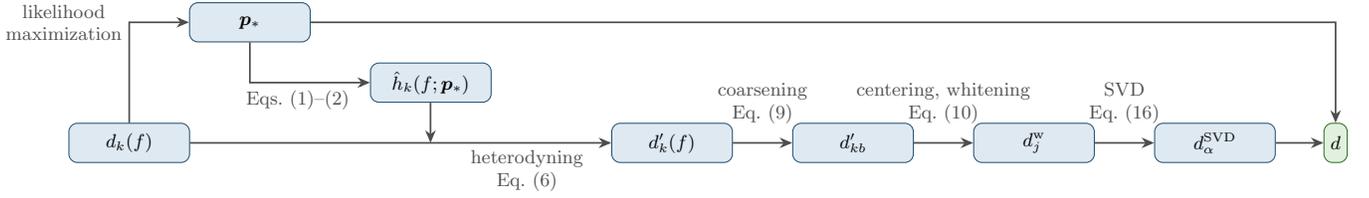

In the remainder of the article, we will exclusively use the compressed representation of the data (i.e., the reference waveform parameters $\bm p_*$ conditioned as described in Appendix~\ref{sec:network-friendly-parametrization}, together with the SVD-compressed heterodyned data $d^\mathrm{SVD}$), and simply refer to it as the data $d$.
The procedure to obtain them is summarized in Fig.~\ref{fig:data_representation}

%%%%%%%%%%%%%%%%%%%%%%%%%%%%%%%%%%%%%%%%%%
\section{Parameter representation}
\label{sec:parameter_representation}
%%%%%%%%%%%%%%%%%%%%%%%%%%%%%%%%%%%%%%%

\begin{figure*}
    \centering
    \begin{minipage}{.33\linewidth}
        \includegraphics[width=0.95\linewidth]{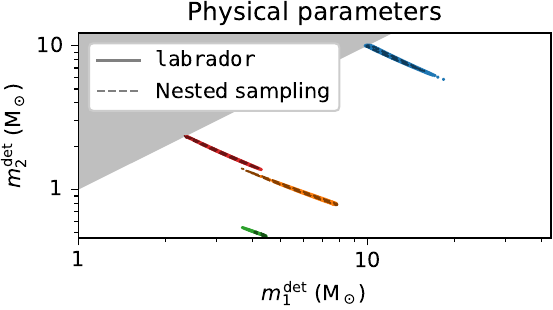}

        \includegraphics[width=0.95\linewidth]{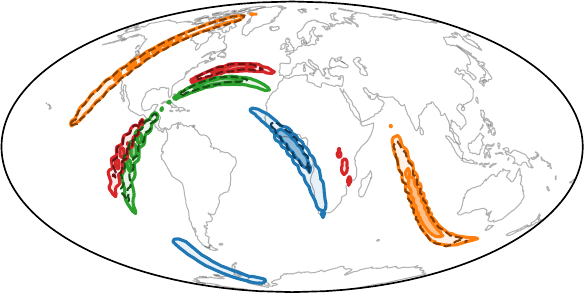}
    \end{minipage}%
    \begin{minipage}{.33\linewidth}
        \includegraphics[width=0.95\linewidth]{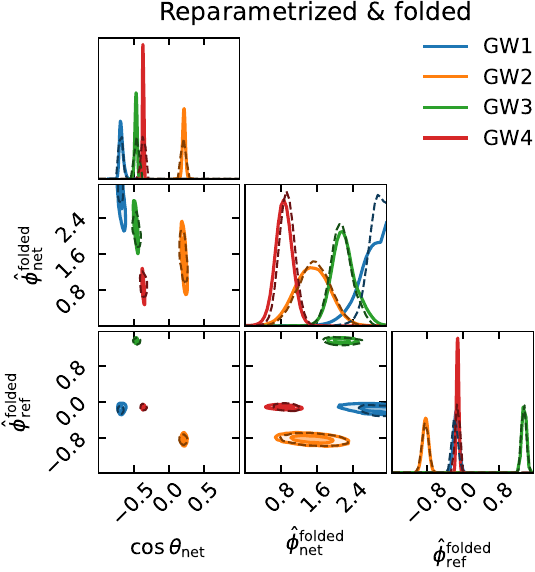}
    \end{minipage}%
    \begin{minipage}{.33\linewidth}
        \includegraphics[width=0.95\linewidth]{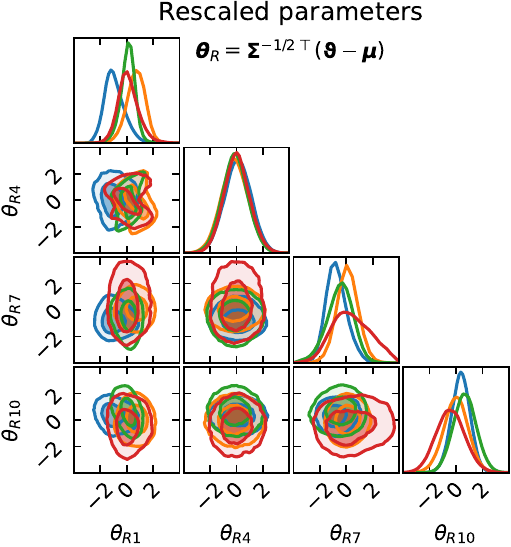}
    \end{minipage}%
    \caption{
        We simplify the parameter representation through a series of transformations.
        \textit{Left:} the physical parameters show nonlinear degeneracy, multimodality, and variability.
        \textit{Center:} we apply an analytical reparametrization to mitigate degeneracy (\S\ref{sec:reparametrization}), and folding to remove multimodality (\S\ref{sec:folding}).
        \textit{Right:} We further rescale these coordinates by a mean and covariance predicted from the data using a small neural network (\S\ref{sec:rescaling}).
        As a result, the posteriors resemble a standard Gaussian and are approximately invariant to the source parameters.
    }
    \label{fig:reparametrizations}
\end{figure*}

Besides the data, the posterior $p(\theta \mid d)$ involves the physical parameters, for which we now aim to construct a representation that is invariant to the source properties.
This may sound paradoxical, but it can be achieved if the parametrization depends on the data.
Consider as an example a normalizing flow: it is a data-dependent coordinate change in which the posterior is a standard multivariate normal, regardless of the properties of the signal---in other words, the source properties are stored in the transformation, not the parameter values.
In this section we aim to ``build a normalizing flow by hand,'' in the sense of devising ourselves the transformation that makes the posterior as close as possible to a standard normal, so that the computationally intensive stage of training the neural posterior estimator gets simplified as much as possible.
We achieve this in three main steps: starting from the usual physical parameters ($\theta$), we apply an analytical transformation to remove nonlinear degeneracies ($\theta_T$), we remove multimodality via folding ($\theta_F$), and rescale the resulting (unimodal, approximately Gaussian) posterior distribution with respect to a mean and covariance predicted from the data by a small neural network ($\theta_R$).
Schematically:
\begin{equation}
    \theta
    \xrightarrow{\text{transform}}
    \theta_T
    \xrightarrow{\text{fold}}
    \theta_F
    \xrightarrow{\text{rescale}}
    \theta_R\,.
\end{equation}
These transformations are illustrated in Fig.~\ref{fig:reparametrizations}, and described in more detail in the following subsections.

%%%%%%%%%%%%%%%%%%%%%%%%%%%%%%%%%%%%%%%%%%%
\subsection{Removing degeneracies via reparametrization}
\label{sec:reparametrization}

Approximate analytic models of the gravitational waveform and how it enters the likelihood function allow us to identify specific parameter combinations that are well constrained, such as the amplitude, phase and time of arrival of the wave at each detector, as well as certain combinations of the coefficients of the post-Newtonian expansion.
In general, these are nonlinear functions of the physical parameters of interest (masses, spins, source location), whose posterior distributions therefore tend to show complex structure.
Building on \citet{Roulet2022} for the extrinsic parameters, and \citet{Lee2022} for the intrinsic, we adopt a parametrization $\theta_T$ that promotes the observable combinations to coordinates, mitigating nonlinear degeneracies.

%---------------------------------------
\subsubsection{Extrinsic parameters}

With the heuristic that the amplitude, phase and time of arrival at each detector are usually well measured, \citet[Table~I]{Roulet2022} recast the extrinsic parameters in terms of the detector-network-based sky location angles $(\theta_\textrm{net}, \hat \phi_\textrm{net})$; and, for a chosen reference detector, the arrival time $t_{k_0}$, phase $\hat\phi_\textrm{ref}$, and inverse-amplitude (or ``chirp distance'' \cite{Brady2008}) $\hat D$.
We adopt this parametrization, with the following modifications.

These coordinates require specifying a reference frequency $f_\textrm{ref}$, a primary and secondary detector $(k_0, k_1)$, a maximum-likelihood arrival phase at the primary detector $\varphi_{k_0}^\textrm{ML}$, and an average frequency $\overline f_{k_0}^\textrm{ML}$.
While in Ref.~\cite{Roulet2022} these quantities are determined separately for each event, here we set most of them globally to ensure homogeneity in the training set.
Specifically, we choose the primary and secondary detectors as the two most sensitive in the network (currently LIGO Livingston and LIGO Hanford), and fix both the reference and average frequencies near the peak of detector sensitivity, $f_\textrm{ref} = \overline f_{k_0}^\textrm{ML} = \SI{100}{\hertz}$.

The maximum-likelihood phase is still obtained on a per-event basis as the argument of the frequency-domain reference waveform at the reference frequency, $\varphi_{k_0}^\textrm{ML} = \Phi_{k_0}(f_\mathrm{ref}; \bm p_\ast)$. 
This ensures that the arrival phase at the primary detector $\hat\phi_\textrm{ref}$ remains close to $0 \pmod{\pi}$ for all events.
In a similar spirit, we parametrize the arrival time and amplitude relative to those of the reference waveform at the primary detector:
\begin{align}
    t_\textrm{rel} &= t_{k_0}(t_\oplus, \los) - t_{k_0*}\,,\\
    \hat D_\textrm{rel}(\mathcal M, D, \iota, \los, \psi)
    &= \frac{A_{k_0*}}{A_{k_0}(\mathcal M, D, \iota, \los, \psi)}\,.\label{eq:d_rel}
\end{align}
Analytical expressions for the predicted time $t_k$ and amplitude $A_k$ in terms of the geocenter time $t_\oplus$, chirp-mass $\mathcal M$, luminosity distance $D$, inclination $\iota$, sky location $\los$ and polarization $\psi$ are found e.g.\ in Ref.~\cite{Roulet2022}.

%---------------------------------------
\subsubsection{Intrinsic parameters}

For low-mass (inspiral-dominated) sources, \citet{Lee2022} introduced coordinates $(\mu_1, \mu_2, q, \chi_{2z})$, where $\mu_1$ and $\mu_2$ are functions of $(\mathcal{M}, q, \chi_{1z}, \chi_{2z})$, with $\mathcal{M}$ the chirp mass, $q$ the mass ratio, and $\chi_{1z}$ and $\chi_{2z}$ the components of the primary and secondary dimensionless spin vectors along the orbital angular momentum.
The quantities $(\mu_1, \mu_2)$ are constructed as the principal components of the Fisher information matrix for the 1.5 PN phase coefficients of the waveform.
The heuristic behind this is that $(\mu_1, \mu_2)$ are typically well-measured from the phase evolution, while $(q, \chi_{2z})$ parametrize degenerate directions that are usually more constrained by the prior than by the likelihood.

Our implementation largely follows these choices, with a modification.
The parameters $(\mu_1, \mu_2, q)$ together determine the value of the two masses, as well as the leading PN spin parameter $\beta$, which is a function of $(q, \chi_{1z}, \chi_{2z})$ \cite{Lee2022}.
Given $q$ and $\beta$, specifying $\chi_{2z}$ also determines $\chi_{1z}$; however, some choices of $\chi_{2z}$ yield unphysical $|\chi_{1z}| > 1$.
This places additional constraints on $\chi_{2z}$ that depend on $(\mu_1, \mu_2, q)$, inducing correlations and irregular boundaries in the posterior.
To avoid this, we replace $\chi_{2z}$ by 
\begin{equation}
    C_{2z} = \frac{
        \chi_{2z}
        - \chi_{2z}^\mathrm{min}
    }{
        \chi_{2z}^\mathrm{max}
        - \chi_{2z}^\mathrm{min}
    }\,,
\end{equation}
where $\chi_{2z}^\mathrm{min}, \chi_{2z}^\mathrm{max}$ are functions of $(\mu_1, \mu_2, q)$ that enforce $|\chi_{1z}|<1$ and $|\chi_{2z}|<1$.
Thus, in the simplifying limit that the secondary spin $\chi_{2z}$ does not affect the likelihood at constant $(\mu_1, \mu_2)$ and has a uniform prior, the conditional posterior of $C_{2z}$ is uniform in $(0, 1)$ irrespective of the data and other source parameters.

For high-mass sources ($\mathcal{M} \gtrsim \SI{15}{M_\odot}$), the post-Newtonian-based parametrization in terms of $(\mu_1, \mu_2)$ is not as justified, since the merger-ringdown contributes a significant amount of information. These variables can even be problematic because they involve negative powers of the chirp mass, which at high mass become singular.
Since mass measurements in this regime are generally less precise and less correlated, we simply use the chirp mass, log mass-ratio, effective spin $\chi_\mathrm{eff}$ and $C_{2z}$.

All these transformations are analytic and invertible.

%%%%%%%%%%%%%%%%%%%%%%%%%%%%%%%%%%%%%%%%%%%%
\subsection{Removing multimodality via folding}
\label{sec:folding}

A network of two co-aligned gravitational-wave detectors is not particularly effective at distinguishing
\begin{enumerate}
    \item a source overhead from one underfoot,
    \item right- from left-handed polarization,
    \item an orbital phase shift of $\pi$,
    \item a simultaneous phase and polarization shift by $\pi/2$;
\end{enumerate}
these approximate discrete symmetries generically give rise to posteriors with up to $2^4=16$ modes (occasionally fewer, for events with high signal-to-noise ratio, three detectors, or measurable higher-order modes).

Following \citet{Roulet2022}, we fold the parameter space along these symmetry directions to produce unimodal posteriors.
For example, the folded cosine-inclination is $(\cos\iota)_F = - |\cos\iota|$, whose posterior distribution typically has a single peak near $-1$, instead of two approximately symmetric ones at $\pm 1$ for right- or left-polarized waves.
Because the folding operation is not invertible, we must recast the parameters $\theta_T$ in terms of the folded coordinates $\theta_F$ and a discrete label $\ell \in\{1, \ldots, 16\}$ that specifies which symmetry sector they should be unfolded into:
\begin{equation}
    \theta_T \leftrightarrow (\theta_F, \ell)\,.
\end{equation}
Being unimodal, the folded distribution $p(\theta_F \mid d)$ can be represented with a simpler model.

To reconstruct the full posterior
\begin{equation}
    p(\theta_T \mid d)
    = p(\theta_F\mid d)\, p(\ell\mid \theta_F, d)\,,
\end{equation}
we additionally train a classifier that predicts $p(\ell\mid \theta_F, d)$.
We use the \texttt{XGBoost} \cite{Chen2016} implementation of gradient-boosted decision trees with a multiclass probabilistic objective (softmax cross-entropy), trained on the same dataset as the flow (in practice, we do not condition the classifier on $\theta_F$, but on its rescaled version $\theta_R$, defined next).
Then, for each sample $\theta_F$ we generate a label $\ell \sim p(\ell \mid \theta_F, d)$ and use it to invert the folding transform.

%%%%%%%%%%%%%%%%%%%%%%%%%%%%%%%%%%%%%%%%%%%%%%%
\subsection{Removing variability via rescaling}
\label{sec:rescaling}

Despite being satisfactorily unimodal, the folded posterior $p(\theta_F \mid d)$ still has some unwanted features:
\begin{enumerate}
    \item Some parameters are periodic, e.g.\ the polarization angle $\psi \in [0, \pi)$.
    If the posterior peaks near $0/\pi$, it can split, producing a spurious bimodality.
    \item Certain parameters have sharp boundaries, e.g.\ $q \leq 1$, requiring more expressive flows.
    \item Naturally, the location and scale of the posterior depend on the data.
\end{enumerate}
To address these issues, we introduce an invertible rescaling transform $\theta_F \leftrightarrow \theta_R$ designed so that, ideally, the rescaled posterior $p(\theta_R \mid d)$ resembles a standard Gaussian.
The transformation has three main stages:

\paragraph{Unconstraining transform.}
Parameters bounded to a finite interval are mapped to $\mathbb{R}$ using the inverse cumulative of the normal distribution (probit function), so that uninformative parameters with a uniform posterior become standard normal.\footnote{We found that probit performs better than the often used logit due to this property.}
For periodic parameters, to avoid bimodality at the boundaries, we first center them by predicting their circular mean from the data (see \textit{c.}\ below), and then map them to $\mathbb{R}$ using the unconstraining transform.

\paragraph{Affine standardization.}
The nonperiodic parameters are linearly decorrelated from the data by subtracting a least-squares affine fit, and their residuals are scaled to have unit variance over the whole training set.
This step ensures that parameters have a $\mathcal{O}(1)$ dynamic range before the neural-network-based whitening.

\paragraph{Neural rescaling.} \label{step:neural}
A neural network predicts the conditional mean $\mu$ and covariance $\Sigma$ of the preconditioned (through steps \textit{a} and \textit{b}) parameters $\vartheta$, given the compressed data.
The parameters are then whitened by removing the predicted mean and scale:
\begin{equation}
    \theta_R = \Sigma^{-1/2\top}(d) \cdot(\vartheta - \mu(d))\,,
\end{equation}
where $\Sigma^{-1/2}$ is the Cholesky factor of $\Sigma^{-1}$.
The network parametrizes $\Sigma^{-1/2}$ by outputting the logarithm of its diagonal elements and the lower-triangular off-diagonal entries, enforcing a positive definite covariance.
The network is trained by minimizing a cross-entropy loss $L(\phi)$ between the empirical distribution of preconditioned parameters and a multivariate normal $\mathcal N$:
\begin{equation}
    L(\phi)
    = -\sum_{i} w_i \log \mathcal N \big(\vartheta_i; \mu_\phi(d_i), \Sigma_\phi(d_i)\big)\,,
\end{equation}
where $\phi$ are the trainable parameters ($w_i$ are importance-sampling weights discussed in \S\ref{sec:weighting}).
To prevent sharp features in the rescaling transform we use a small multilayer perceptron (say, 4 layers of 32 neurons) with a smooth activation function (Sigmoid Linear Unit, SiLU). The small size of the network also keeps the cost of training the rescaler low ($\SI{1}{GPU\,h}$ on an NVIDIA V100 for \num{e7} simulations).

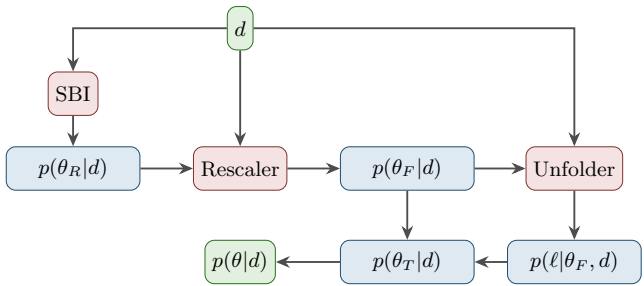
\begin{figure}
    \centering
    \def\hstep{2.5}
\def\vstep{0.7}

\begin{tikzpicture}[
    bluebox/.style={
    rectangle, rounded corners=4pt,
    draw=bluedraw, fill=bluefill,
    minimum width=2.0cm, minimum height=0.65cm,
    align=center, font=\small
    },
    redbox/.style={
    rectangle, rounded corners=4pt,
    draw=reddraw, fill=redfill,
    minimum height=0.65cm,
    align=center, font=\small
    },
    greenbox/.style={
    rectangle, rounded corners=4pt,
    draw=greendraw, fill=greenfill,
    minimum height=0.65cm,
    align=center, font=\small
    },
    arr/.style={-{Stealth[length=6pt, width=5pt]}, thick, color=black!70},
    darr/.style={-{Stealth[length=6pt, width=5pt]}, semithick, dashed, color=black!50},
]
% Bottom row
\node[bluebox] (thetaR) at (0,0)    {$p(\theta_R|d)$};
\node[redbox] (rescaler)  at (\hstep,0)  {Rescaler};
\node[bluebox] (thetaF) at (2*\hstep,0)  {$p(\theta_F|d)$};
\node[bluebox] (ell)  at (3*\hstep,-2.0*\vstep)  {$p(\ell|\theta_F, d)$};
\node[bluebox] (thetaT) at (2*\hstep,-2.0*\vstep) {$p(\theta_T|d)$};
\node[greenbox] (theta)    at (\hstep,-2.0*\vstep) {$p(\theta|d)$};

% Middle row
\node[redbox] (sbi)  at (0.0,1.6*\vstep)  {SBI};
\node[redbox] (xgb)  at (3*\hstep,0.0*\vstep)  {Unfolder};

% Top row
\node[greenbox] (d)  at (\hstep,3.0*\vstep)  {$d$};

% --- Arrows ---
\draw[arr] (thetaR) -- (rescaler);
\draw[arr] (rescaler) -- (thetaF);
%\draw[arr] (thetaF) -- (ell);
\draw[arr] (ell) -- (thetaT);
\draw[arr] (thetaT) -- (theta);
\draw[arr] (d) -| (sbi);
\draw[arr] (d) -| (xgb);
\draw[arr] (d) -- (rescaler);
\draw[arr] (sbi.south) -| (thetaR);
\draw[arr] (xgb.south) -| (ell);
\draw[arr] (thetaF.east) |- (xgb);
\draw[arr] (thetaF) -- (thetaT);

\end{tikzpicture}
    \caption{Inference workflow with \texttt{labrador}.
    Conditioned on the data $d$, rescaled-parameter samples $\{\theta_R\}$ are generated via neural posterior inference using \texttt{sbi} \cite{Boelts2025}.
    Then, the transformations of Fig.~\ref{fig:reparametrizations} are inverted to reconstruct the physical parameters: the rescaling is undone to obtain folded parameters $\{\theta_F\}$, and each sample is unfolded by generating an unfolding $\ell^i \sim p(\ell \mid \theta_F^i, d)$. 
    }
    \label{fig:labrador_flowchart}
\end{figure}

The result of this sequence of bijective transformations is a coordinate system in which the posterior distribution is approximately standard normal.
Finally, we train a neural posterior estimator to predict this distribution, using the software package \texttt{sbi} \cite{Boelts2025}.
Appendix~\ref{app:architecture} gives technical details.
At inference time, we generate samples from the neural posterior and apply the inverse transformations in reverse order, as outlined in Fig.~\ref{fig:labrador_flowchart}.

%%%%%%%%%%%%%%%%%%%%%%%%%%%%%%%%%%%
\section{Training set}
\label{sec:training_set}
%%%%%%%%%%%%%%%%%%%%%%%%%%%%%%%%%%

In this section we describe the generation of a low-discrepancy training set and introduce a novel, general method to reweight the simulation prior to the inference prior during training in a numerically stable manner.

\subsection{Simulations}

Simulating the training set from a physically motivated prior is undesirable, as it generally induces an inhomogeneous sampling of the data space.
For instance, under a uniform distribution in volume most signals would be far away and have low amplitude.
Instead, it is generally better to simulate from a proposal that more uniformly covers the range of signal morphologies---much like the density of templates used in gravitational-wave searches differs from the astrophysical prior.
With this motivation we adopt the proposal of Table~\ref{tab:simulation_prior}.
We only keep simulations for which the signal-to-noise ratio (SNR) of the reference waveform $h(\bm{p_*})$ is between 8 and 50, so as to focus on the regime relevant to the large majority of the detections. Therefore, the inference model is valid only in this range.

\begin{table}
    \centering
    \caption{Distribution of simulations used in this work; it differs from the inference prior in the luminosity distance and masses, which we choose uniform in luminosity volume and detector-frame masses.
    Besides these ranges, we apply a cut on the network signal-to-noise-ratio $8 < \langle h(\bm{p_*}) \mid h(\bm{p_*}) \rangle^{1/2} < 50$ by dropping simulations outside this range from the training set.
    Arrival time is inferred along with these parameters.
    }
    \label{tab:simulation_prior}
    \begin{ruledtabular}
        \begin{tabular}{ll}
             Parameter & Simulation distribution \\
             \hline
             \rule{0pt}{10pt}%
             $\hat D$ & $\mathrm{U}(\hat D_{50}, \hat D_{8})$\footnotemark[1] \\
             $\mathcal M$ ($\mathrm{M}_\odot$) & $\text{log-uniform}\lowmassrange \text{ or } \highmassrange$ \\
             $q$ & $\text{log-uniform}\qrange$ \\
             $\chi_\mathrm{eff}, \cos\iota, \sin \delta$ & $\mathrm{U}(-1, 1)$ \\
             $C_{2z}$ & $\mathrm{U}(0, 1)$ \\
             $\phi_\mathrm{ref}, \alpha$ & $\mathrm{U}(0, 2 \pi)$ \\
             $\psi$ & $\mathrm{U}(0, \pi)$ \\
        \end{tabular}
    \end{ruledtabular}
    \footnotetext[1]{
        $\hat D_\varrho \coloneqq \num{3e4}
            (\mathcal{M}/\mathrm{M}_\odot)^{-5/6}
            |R_\rml(\los, \iota, \psi)|^{-1}
            (D / \mathrm{Mpc}) \varrho^{-1}
        $ is intended to approximate the chirp distance $\hat D$ \cite{Brady2008,Roulet2022} at which the signal has an expected signal-to-noise ratio $\varrho$ at the Livingston detector.
    }
\end{table}

Taking advantage of the separable and analytical form of this distribution, we generate the training set on a scrambled Halton sequence \cite{Halton1960, Owen2017} using inverse transform sampling.
This turns the Monte Carlo approximation to the cross-entropy loss (see Eq.~\eqref{eq:loss_mc} below) into a more accurate quasi-Monte Carlo estimate.
To preserve the low-discrepancy property of the sequence at the batch level, we implement modifications to the \texttt{sbi} code so that batches consist of contiguous samples instead of a random reshuffling at every epoch.
Similar to Refs.~\cite{Lye2020, Mishra2021}, we find that using a quasirandom sequence reduces overfitting compared to a random training set of the same size.
For the validation set we still use independent samples, to prevent spurious correlations with the training set.

%%%%%%%%%%%%%%%%%%%%%%%%%%%%%%%%%%%%%%%%%%%%%%%
\subsection{Reweighting the simulation prior}
\label{sec:weighting}

As motivated above, to reduce redundancy in the training set it is desirable to simulate from a proposal that homogeneously covers the data space (optimally the Jeffreys prior \cite{Jeffreys1946}, although it is generally intractable), rather than a physically motivated prior.
It is possible to do amortized inference under mismatched simulation and inference priors by reweighting the posterior samples in post-processing \cite{Papamakarios2016, Green2021}.
In this section, we introduce a novel loss function which instead reweights the training set in a controlled way, allowing us to train the model with arbitrarily distributed simulations and directly produce posteriors under the physical prior.

In general, a conditional estimator $q_\phi(\theta \mid d)$ can be trained to approximate the posterior $p(\theta \mid d)$ by minimizing a loss function of the form
\begin{equation}
    L(\phi) =
        -\int\rmd d\, \nu(d)
        \int \rmd \theta \,
        p(\theta \mid d) \log q_\phi(\theta \mid d)\,,
    \label{eq:loss_general}
\end{equation}
with $\nu(d) > 0$ \cite{Papamakarios2016}.
Indeed, the $\theta$ integral is the cross-entropy between the target posterior and the estimator, which is minimized when the two distributions are equal (for the corresponding $d$).
To make them equal for any $d$, the cross entropy is integrated with respect to a positive measure $\nu(d) \rmd d$ that one is free to choose.

Let us parametrize the measure as $\nu(d) = \rme^{-\lambda (d)} p(d)$, which allows Monte Carlo estimation of the loss function using simulations:
\begin{equation}
    L(\phi)
    \approx -\frac{1}{N} 
    \sum^N_{
        \subalign{
            \theta_i &\sim \tilde p(\theta) \\
            d_i &\sim p(d \mid \theta_i)
        }
    }\hspace{-5pt}
    \rme^{-\lambda(d_i)}
    \frac{p(\theta_i)}{\tilde p(\theta_i)}
    \log q_\phi(\theta_i \mid d_i)\,.
    \label{eq:loss_mc}
\end{equation}
Because we draw simulations from a proposal $\tilde p(\theta)$, we pick up importance weights $p(\theta) / \tilde p(\theta)$ (similarly to the so-called SNPE-B method \cite{Lueckmann2017}).
These weights can differ by many orders of magnitude across simulations, which effectively reduces the size of the training set and increases the variance of the (quasi-)Monte Carlo estimator of Eq.~\eqref{eq:loss_mc} (this is the main limitation of SNPE-B \cite{Greenberg2019}).
To mitigate this variance, we design $\lambda(d)$ so as to counteract the imbalance of the importance weights. With this heuristic we choose
\begin{equation}\label{eq:lambda}
    \lambda(d)
    = \alpha_1 \mu(d) + \alpha_2 \sigma(d)\,,
\end{equation}
where $\alpha_1, \alpha_2$ are tunable coefficients and
\begin{equation}\label{eq:mu_sigma}
    \begin{split}
        \mu(d) &= \mathbb{E}_{
            \theta \sim \tilde p(\theta \mid d)
        } 
        \left[
            \log \frac{p(\theta)}{\tilde{p}(\theta)}
        \right]\,, \\
        \sigma^2(d)
        &= \mathbb{E}_{
            \theta \sim \tilde p(\theta \mid d)
        } 
        \left[
            \log^2 \frac{p(\theta)}{\tilde{p}(\theta)}
            - \mu^2(d)
        \right]
    \end{split}
\end{equation}
are regressors for the mean and variance of the log importance-weights, with $\tilde p(\theta \mid d) \propto \tilde p(\theta) \,p(d \mid \theta)$.
We build the regressors of Eq.~\eqref{eq:mu_sigma} with \texttt{XGBoost} \cite{Chen2016} trained on the same supervised learning pairs $(d_i, \theta_i)$ of Eq.~\eqref{eq:loss_mc}; this training takes only few seconds.
Then, we tune the coefficients by maximizing the effective sample size of the training set numerically:
\begin{align}
    n_\mathrm{eff}(\alpha_1, \alpha_2)
    &= \frac{
        (\sum_i w_i)^2
    }{
        \sum_i w_i^2
    }\,;\\
    w_i
    &= \rme^{-\lambda(d_i; \alpha_1, \alpha_2)}
    \frac{p(\theta_i)}{\tilde p(\theta_i)}\,. \label{eq:prior_weights}
\end{align}
The form of $\lambda(d)$ chosen in Eq.~\eqref{eq:lambda} attempts to improve the reweighting efficiency by balancing the weights in Eq.~\eqref{eq:prior_weights}.
We introduce the regressors $\mu(d), \sigma(d)$ because we need a function of the data, $\rme^{-\lambda(d)}$, to counterweight a function of the parameters, $p(\theta) / \tilde p(\theta)$.
The first term in Eq.~\eqref{eq:lambda} counterbalances the expected prior ratio, which is achieved with $\alpha_1 \approx 1$, and the second downweights (for $\alpha_2 > 0$) large deviations whenever the expectation is uncertain.

As per Table~\ref{tab:simulation_prior}, we simulate signals according to a proposal distribution that is uniform in inverse signal-to-noise ratio, and log-uniform in detector-frame chirp mass and mass-ratio.
For inference, we instead use a prior uniform in luminosity volume and detector-frame masses.
Directly reweighting the training set to our inference prior, i.e., setting $\lambda(d) = 0$, would have a low efficiency $n_\textrm{eff} / N = \effrawweights$.
In contrast, our choice of $\lambda(d)$ improves the efficiency to $\effcounterweights$, about $\effratio\times$ higher.
We find that including additional tunable coefficients and moments in Eq.~\eqref{eq:lambda} yields diminishing returns in our case study.

The ``counterweight'' method we introduced here is generally applicable beyond the gravitational-waves domain, and in particular it would be interesting to further explore its use in sequential neural posterior estimation---an inference technique in which the simulations are, by design, not drawn from the physical prior.
It is worth mentioning the automatic posterior transformation (SNPE-C) \cite{Greenberg2019}, another algorithm that achieves inference with an arbitrary simulation prior by modifying the loss function. Unlike our method, though, SNPE-C has the drawback that the loss function does not penalize posterior leakage into regions with no prior support, requiring further mitigation techniques \cite{Deistler2022, Xiong2025}.

%%%%%%%%%%%%%%%%%%%%%%%%%%%%%%%%%%%%%%%
\section{Convergence and performance}
\label{sec:convergence}
%%%%%%%%%%%%%%%%%%%%%%%%%%%%%%%%%%%%%%

\begin{figure}
    \centering
    \includegraphics[width=.7\linewidth]{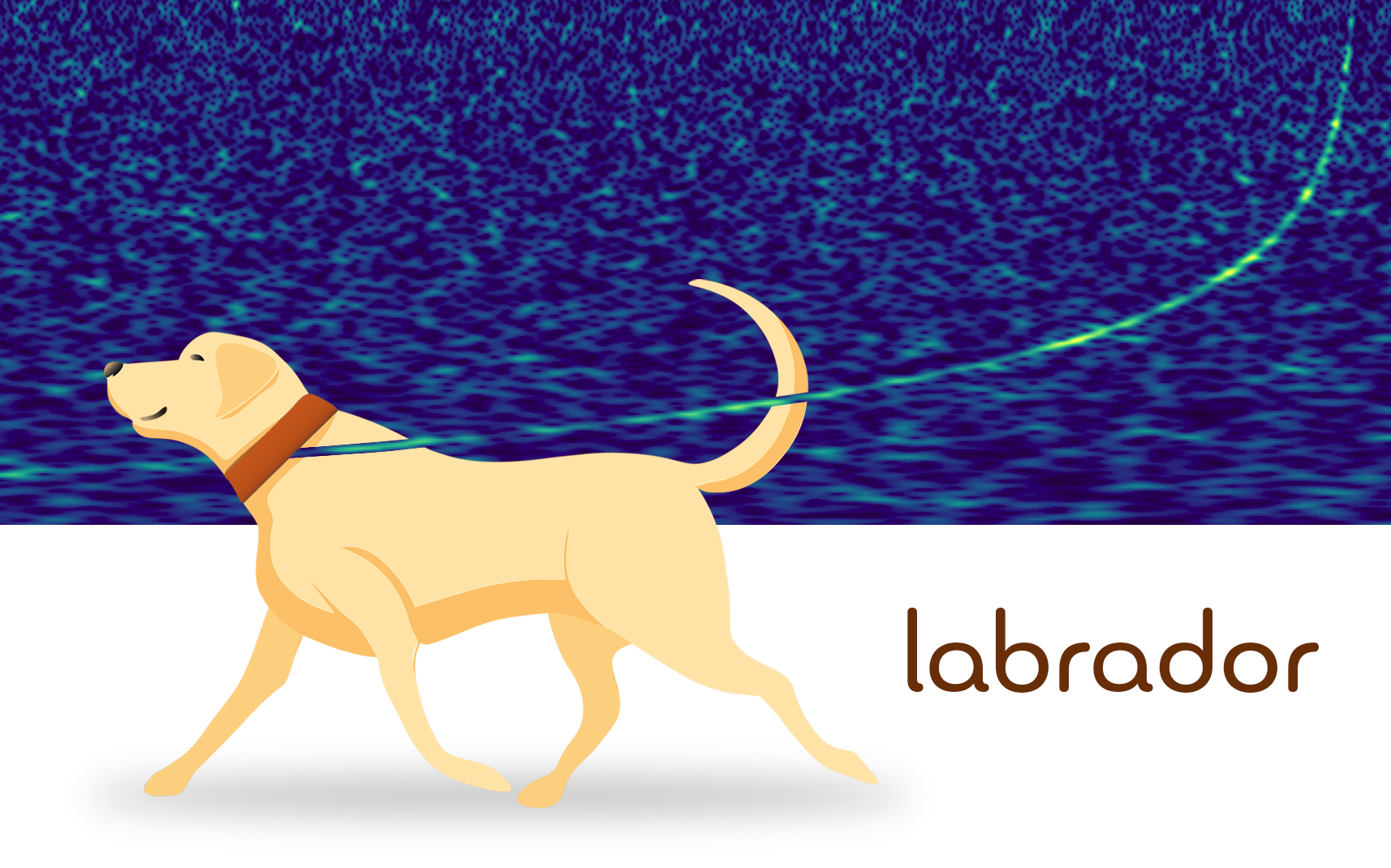}
    \caption{Logo of \texttt{labrador}.
    }
    \label{fig:labrador}
\end{figure}

We implement the methods described above in an open-source code named \texttt{labrador}, Fig.~\ref{fig:labrador},\textsuperscript{\ref{ftn:github}}
built on top of the gravitational-wave inference software \texttt{cogwheel} \cite{Roulet2022}.
In this section we validate its performance on simulations and on LIGO data from the O4a observing run \cite{Abac2026,Abac2026b}.
We model gravitational-wave signals using the aligned-spin, quadrupole-radiation \texttt{IMRPhenomXAS} approximant \cite{Pratten2020}
(recall that inference follows the simulation model, even though we use a basic phenomenological model for compressing the data).
As reported in Table~\ref{tab:simulation_prior}, we train two models, for low- ($\mathcal{M} \in 1$--$\SI{15}{M_\odot}$) and high-mass ($\mathcal{M} \in 12$--$\SI{50}{M_\odot}$) sources.
We make the trained models available at \cite{Roulet2026}.

%%%%%%%%%%%%%%%%%%%%%%%%%%%%%%%%%%%%%%%
\subsection{Computational cost}

\begin{figure}
    \centering
    \includegraphics[width=\linewidth]{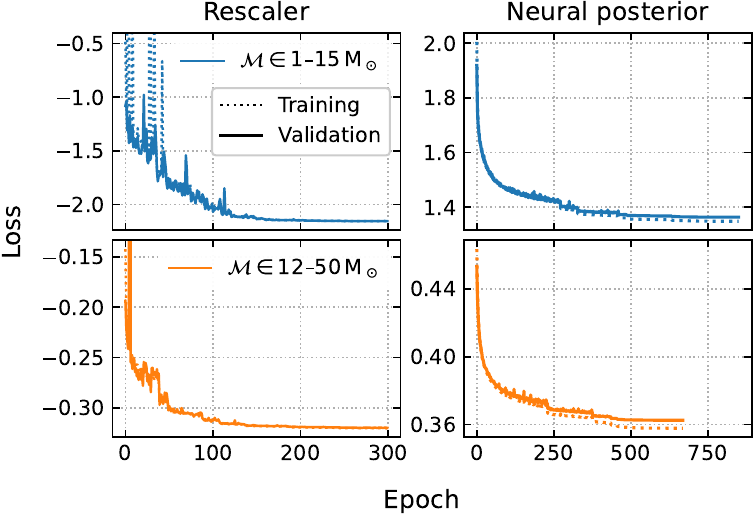}
    \caption{Loss function as a function of training epoch for the rescaler (left, \S\ref{sec:rescaling}) and the neural posterior estimator (right), for our low-mass (top) and high-mass (bottom) models. The neural estimator shows some overfitting.
    }
    \label{fig:loss}
\end{figure}

The two steps that dominate the computational cost of training a \texttt{labrador} model are generating the training set and training the \texttt{sbi} posterior estimator---in this work, a conditional normalizing flow.
With our current implementation, each simulation takes \SI{\secpersimulation}{\second} on average (dominated by finding the best-fit reference waveform), bringing the cost of generating \num{e7} simulations to {\cpudays} CPU core-days.
Thus, access to at least $\mathcal{O}(100)$ cores is preferable for practicality (we have implemented parallelization by means of the \texttt{HTCondor} scheduler \cite{Thain2005}). 
The training step is done separately and takes \SI{10}{h} on an NVIDIA Tesla V100 GPU, with the model architecture described in Appendix~\ref{app:architecture}.
Additional steps include training the rescaler (\SI{1}{GPU}-h, see \S\ref{sec:rescaling}); and compressing the data (\S\ref{sec:compression}), computing the prior weights (\S\ref{sec:weighting}), and training the unfolding classifier (\S\ref{sec:folding}), all under $\SI{1}{CPU}$-h.
Figure~\ref{fig:loss} shows the loss function during the training of these models.
Some overfitting is visible in the neural posterior; we find that it worsens if we use larger neural networks.

After the model has been trained, \texttt{labrador} generates over {\samplespersec} samples/s on a single CPU core for a given event.
If desired, the accuracy of the result can be improved with importance sampling (similar to \citet{Dax2023}), applying a weight to each sample according to \cite{Ashton2026}
\begin{equation}
    \label{eq:is_weights}
    w_i = \frac{p(\theta_i) \,\mathcal L(d \mid \theta_i)}{q(\theta_i \mid d)}\,,
    \quad\theta_i \sim q(\theta \mid d)\,,
\end{equation}
where $q(\theta \mid d)$ denotes the \texttt{labrador} posterior from which samples are drawn, $p(\theta)$ the inference prior described in \S\ref{sec:weighting} and $\mathcal L (d \mid \theta)$ the likelihood function \eqref{eq:likelihood}.
This step corrects an imperfectly learned posterior; however, in this case the inference is no longer amortized and the computational cost becomes dominated by evaluations of the posterior, for which we use \texttt{cogwheel}.
With the \texttt{IMRPhenomXAS} waveform model, we produce $\issamplespersec$ weighted samples/s on one CPU core.
Because of the uneven weights, the effective sample size $n_\mathrm{eff}$ increases at a lower rate, characterized by the importance-sampling efficiency
\begin{equation}
    \label{eq:is_efficiency}
    \epsilon \coloneqq \frac{n_\mathrm{eff}}{N} = \frac{\overline w^2}{\overline{w^2}}\leq 1\,.
\end{equation}
As we report below, $\epsilon \sim \medianeff$ in the current \texttt{labrador} implementation.
Importance sampling additionally yields the Bayesian evidence
\begin{equation}
    \mathcal Z
    = \int p(\theta) \mathcal L (d \mid \theta)\,\rmd \theta \approx \overline w \,,
\end{equation}
which has applications in model selection and signal detection, and the effective sample size $n_\mathrm{eff}$, which constitutes a very powerful convergence test.

\subsection{Performance on injections}

In this section we diagnose the performance of \texttt{labrador} on the validation set, consisting of synthetic injections in Gaussian, stationary noise with a fixed power spectrum characteristic of the O4a observing run on the two LIGO detectors \cite{Abac2025}.

\begin{figure}
    \centering
    \includegraphics[width=\linewidth]{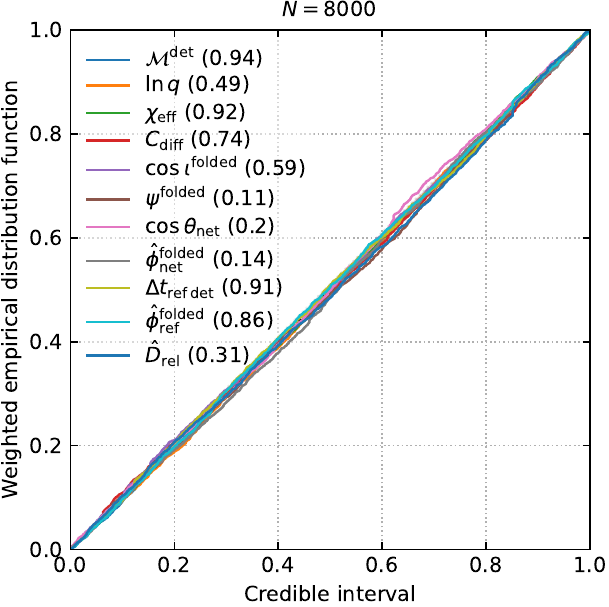}
    \caption{
        Probability--probability plot for parameter inference on the validation set, showing the empirical cumulative distribution of the credible interval at which the true parameters are recovered.
        The distribution is uniform, indicating good performance.
        Each simulation is weighted according to Eq.~\eqref{eq:prior_weights} to account for the difference between the prior used to build the training and validation sets and that used for inference.
        $p$-values of (weighted) Kolmogorov--Smirnov tests for uniformity are shown in parentheses; no outliers are detected.
        These results are for the high-mass model in Table~\ref{tab:simulation_prior}; the low-mass model's are similar.
    }
    \label{fig:pp_plot}
\end{figure}

In Fig.~\ref{fig:pp_plot} we confirm with a probability--probability plot that the credible levels of all sampled parameters are uniformly distributed to an excellent degree.
The credible intervals correspond to the amortized posteriors $q(\theta \mid d)$, i.e., without applying the importance-sampling correction of Eq.~\eqref{eq:is_weights}.
Usually, probability--probability plots would use the same prior to draw simulations and do inference.
However, as discussed in \S\ref{sec:training_set} our training and validation sets do not follow the inference prior.
We account for this by weighting each simulation according to Eq.~\eqref{eq:prior_weights} in the empirical distribution function ($y$-axis).
We assess uniformity with a  Kolmogorov--Smirnov (KS) test, whose statistic is the maximum departure of the empirical distribution from the predicted one.
Its $p$-value depends on the KS statistic and the sample size.
Because our samples are weighted, we compute the KS statistic using the weighted empirical distribution, and compute the $p$-value using the effective sample size.

\begin{figure}
    \centering
    \includegraphics[width=\linewidth]{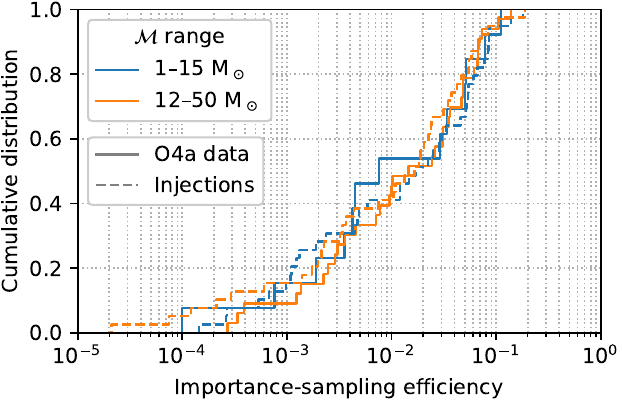}
    \caption{Importance-sampling efficiency ($\epsilon=n_\mathrm{eff} / N$; Eq.~\eqref{eq:is_efficiency}) over a set of simulated (dashed) and real (solid) signals for two \texttt{labrador} models, low mass and high mass. In all cases the median efficiency is about \medianeff, with a broad distribution.
    }
    \label{fig:efficiency}
\end{figure}

As a more stringent test of accuracy of the posterior, in Fig.~\ref{fig:efficiency} we report the efficiency of the importance sampling procedure in Eq.~\eqref{eq:is_efficiency}.
\texttt{labrador} achieves a median efficiency of $\medianeff$, which means that it takes a median of {\mediantimetothousandneff} to reach $n_\textrm{eff} \approx 1000$ with one core. However, the distribution is broad; for example, a few percent of the examples achieve efficiencies below $\num{e-4}$, or above 10\%.

\begin{figure*}
    \centering
    \includegraphics[width=\linewidth]{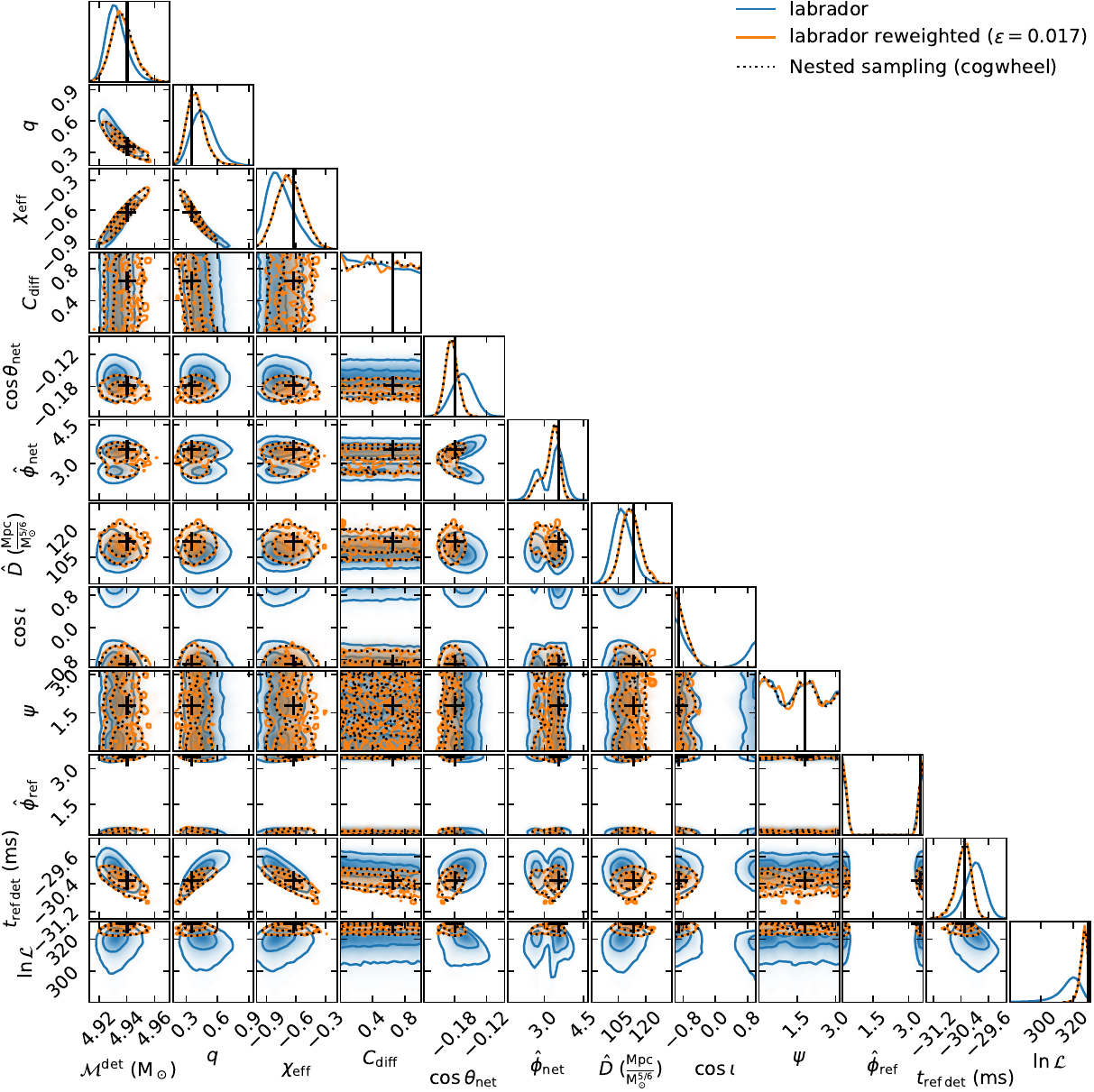}
    \caption{
        Parameter estimation on a simulated signal.
        \texttt{labrador} (this work, solid blue) is about {\labradorfaster} faster than nested sampling via \texttt{cogwheel} \cite{Roulet2022} (dotted black) and gives a good first approximation to the posterior.
        The \texttt{labrador} posterior may be reweighted (dashed orange) using the log-likelihood (last row); this procedure achieves an excellent match to the ground truth nested sampling.
        We chose this simulation as the one with median importance-sampling efficiency (1.7\%).
        The simulated value is shown in black.
        Contours enclose 50\% or 90\% of the distribution.
        See Ref~\cite{Roulet2022} for the definition of the coordinates shown.
    }
    \label{fig:injection_corner_plot}
\end{figure*}

In Fig.~\ref{fig:injection_corner_plot} we show a typical inference with \texttt{labrador} on a simulated signal, and verify that importance sampling brings it to perfect agreement with classic nested sampling (performed using \texttt{cogwheel} \cite{Roulet2022} coupled to the \texttt{nautilus} sampler \cite{Lange2023}).
Note that we use \texttt{cogwheel} to evaluate the posterior for both reweighting \texttt{labrador} and nested sampling, so this is an internal consistency check and a validation of the nested sampler rather than a fully independent comparison.

%%%%%%%%%%%%%%%%%%%%%%%%%%%%%%%%%%%%%%%%
\subsection{Performance on real data}

\begin{figure*}
    \centering
    \includegraphics[width=\linewidth]{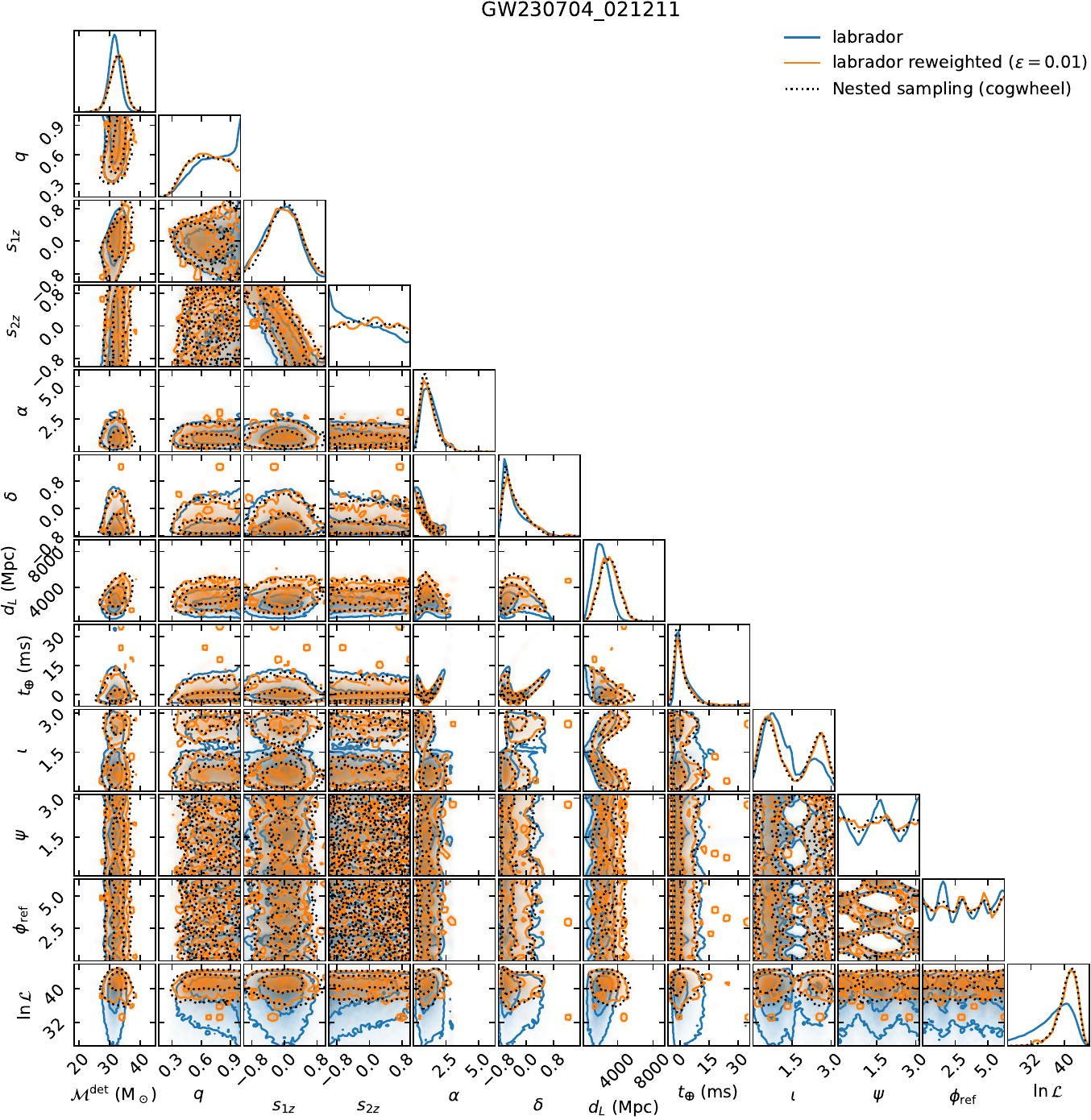}
    \caption{
        Parameter estimation on the real event GW230704\_021211 \cite{Abac2026}.
        Similar to Fig.~\ref{fig:injection_corner_plot}, \texttt{labrador} (solid blue) gives a fast approximation to the posterior that may be reweighted (dashed orange) using the log-likelihood (last row) to match the nested sampling result (dotted black).
        This event has the median importance-sampling efficiency (1\%).
        Contours enclose 50\% or 90\% of the distribution.
    }
    \label{fig:real_event_corner_plot}
\end{figure*}

Lastly, we illustrate the performance of \texttt{labrador} on actual LIGO data.
This requires an additional consideration: a current limitation of \texttt{labrador} is that it is trained with a fixed noise PSD, while in reality the detector characteristics are variable.
A comprehensive solution is to condition the inference on the measured PSD \cite{Wildberger2023}; in lieu, we implement an approximate correction post-training while ignoring PSD uncertainty~\cite{Biscoveanu2020,Plunkett2022}, as follows.
The gravitational-wave signal is the expected value of the measured strain, and the PSD its variance.
The whitening procedure in Eq.~\eqref{eq:heterodyned_data} has the dual purpose of conditioning the data for compression (removing loud narrowband lines) and standardizing the variance.
Standardizing the variance affects the expected value of the whitened data, introducing a bias at linear order in PSD variations.
To avoid this, for real data we implement Eq.~\eqref{eq:heterodyned_data} using a smoothly modified PSD
\begin{equation}
    \begin{split}
        S^{-1}_k(f) &\mapsto S^{-1}_k(f) \, r_k(f)\,, \\
        r_k(f_b) &= \frac{
            \int \rmd f \, f^{-7/3} \, S_{k}^{(0)-1}(f) \, s_b(f)
        }{
            \int \rmd f \, f^{-7/3} \, S^{-1}_k(f) \, s_b(f)
        }\,,
    \end{split}
\end{equation}
where $S_k$ is the actual PSD of the data, $S_k^{(0)}$ is the PSD used in training, and $r_k(f)$ is an interpolating spline with coarse-frequency nodes $\{f_b\}$.
This ensures that we filter the data with an effective PSD that resembles the training one at coarse resolution (hence removing the linear bias at the expense of foregoing variance standardization), while preserving the sharp spectral features of the actual data (hence doing an appropriate conditioning).
We show a representative example of inference in Fig.~\ref{fig:real_event_corner_plot}, this time with our high-mass \texttt{labrador} model.
We observe similar performance in real data and in simulations.
Some outlying samples are apparent in the reweighted posterior for right ascension, declination and geocenter time; they are a manifestation of the relatively small effective sample size, which appears in these variables because their posteriors feature an extended, low-density tail.

\begin{figure}
    \centering
    \includegraphics[width=\linewidth]{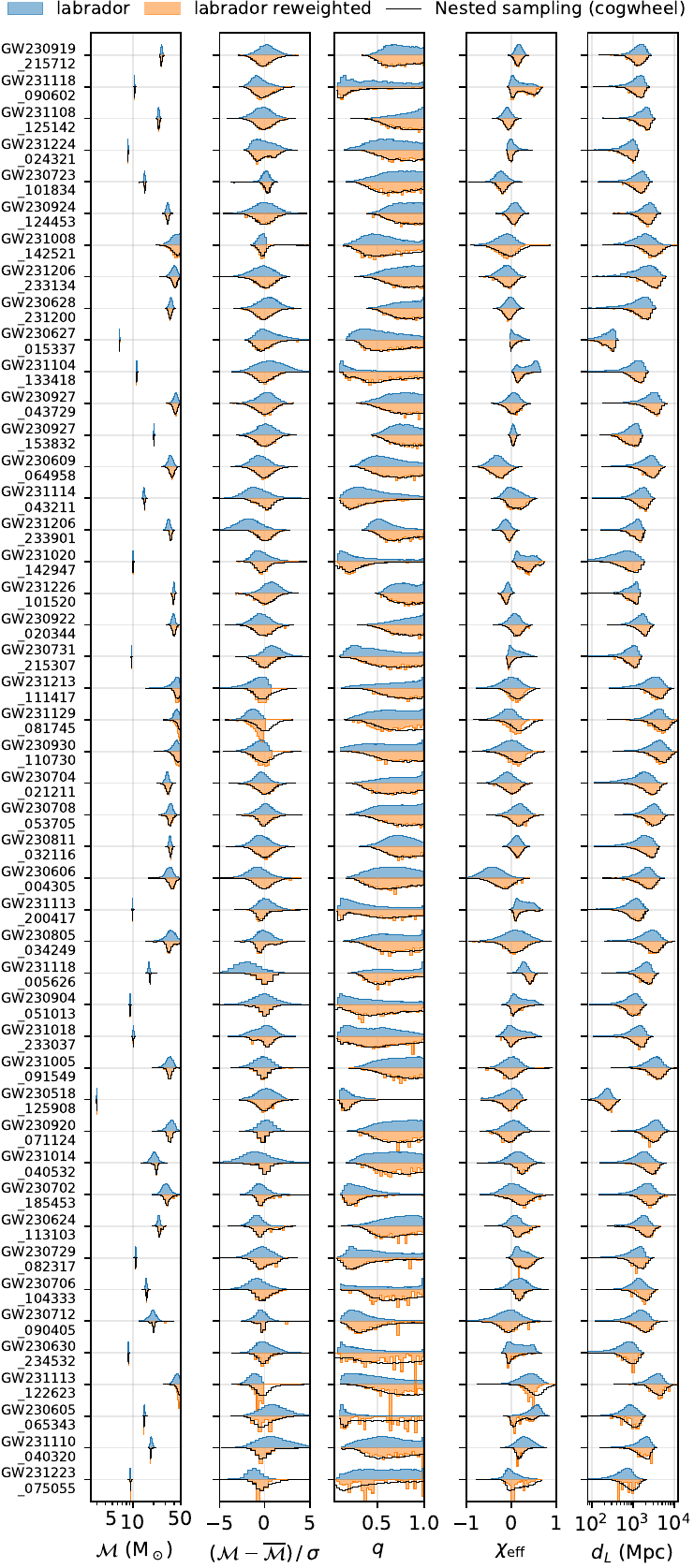}
    \caption{
    Inference on all O4a events from GWTC-4.0 with $\mathcal M \in (1, 50)\,\mathrm{M}_\odot$ and Hanford and Livingston data available.
    Events are sorted by decreasing importance sampling efficiency (Eq.~\eqref{eq:is_efficiency}).
    We show the marginal posteriors on some parameters of interest %, namely chirp mass $\mathcal M$, mass ratio $q$, effective spin $\chi_\mathrm{eff}$ and luminosity distance $d_L$, 
    obtained with \texttt{labrador} (with and without importance-sampling correction, in blue and orange, respectively), and nested sampling (black).
    The 1D posteriors are generally adequate, and after the importance sampling correction they show excellent agreement with nested sampling, modulo noisy estimates for the lowest-efficiency cases.
    }
    \label{fig:all_events}
\end{figure}

As a final diagnostic, in Fig.~\ref{fig:all_events} we study the performance of \texttt{labrador} on all O4a events in the domain of validity of our trained models.
We show posteriors on parameters of interest, namely chirp mass $\mathcal M$, mass ratio $q$, effective spin $\chi_\mathrm{eff}$ and luminosity distance $d_L$.
Because the chirp mass has a considerable dynamic range, in the second column we rescale it by an event-specific characteristic location $\overline{\mathcal{M}}$ and scale $\sigma$ for visualization purposes.
For each event, we show three sets of results: with amortized \texttt{labrador}, after applying an importance sampling correction, and via nested sampling.
The number of \texttt{labrador} samples drawn was such that $n_\mathrm{eff} \approx 10^3$, capped at $N \leq 10^5$.
The conclusions are similar to those from the examples in Figs.~\ref{fig:injection_corner_plot} and \ref{fig:real_event_corner_plot}: the amortized posterior provides a reasonable first approximation and the importance-sampling-corrected result has excellent agreement with nested sampling, although with very noisy density estimates for the $\sim 10 \%$ of events with lowest importance-sampling efficiency.
In a few events the posterior rails against the training boundary $\mathcal M = \SI{50}{M_\odot}$, which can reflect on the marginal posterior of other correlated parameters.
Interestingly, we do not identify any salient trends between the efficiency and the parameter values, or the SNR.
It is reassuring that, even for the worst-performing events, the amortized posterior generally lies in the correct ballpark.

%%%%%%%%%%%%%%%%%%%%%%%%%%%%%%%%%%%
\section{Discussion}
\label{sec:conclusions}
%%%%%%%%%%%%%%%%%%%%%%%%%%%%%%%%%%

In this work we have developed \texttt{labrador}, an amortized simulation-based inference tool for gravitational-wave inference, designed to reduce the computational demands of training machine-learning models by incorporating physical insights and analysis methods from the gravitational-waves domain, together with other generically applicable techniques.

We devised a representation of the data based on heterodyning the gravitational-wave strain against a reference waveform which, unlike previous methods, we obtain by likelihood optimization in multiple variables.
These include two intrinsic parameters (principal components of the waveform phase) as well as the overall time, phase and amplitude of the signal in each of the available detectors.
We facilitate convergence by optimizing the detector phases and amplitudes analytically, and orthonormalizing the remaining coordinates.
The data are represented by the parameters of the reference waveform and the heterodyned strain, which we further compress via SVD.
The efficiency of the compression can be as high as \num{e4} for binary neutron star signals.

Moreover, we have identified a conceptual connection between heterodyning and group equivariance: 
Heterodyned data are approximately invariant to any transformations to the signal that are captured by the tunable parameters of the reference waveform (the invariance is exact if the transformation is also a symmetry of the noise distribution, such as for time or phase shifts).
Meanwhile, the reference-waveform parameters transform in a smooth, straightforward way, which eases interpolation and interpretability.
Accordingly, we obtain a naturally group-equivariant model without a need for a dedicated architecture, and with a much larger class of equivariances than previously accomplished.

We also adapted the specialized coordinates and folding technique of \citet{Roulet2022} to exploit foreknowledge of the characteristic structure of the posterior, including the shape of degeneracies and multimodalities.
These transformations typically make the posterior unimodal and approximately Gaussian, and thus expressible with a simpler model.
We implemented the unfolding step, in which the physical multimodality is reconstructed, via a probabilistic classifier conditional on the data and the folded parameters.

Beyond adapting domain-specific techniques to neural inference, we also designed novel methods that advance machine learning more broadly.
We developed a technique that enables the use of different priors for simulation and inference, by reweighting the training set a priori instead of the samples a posteriori.
Numerical stability is ensured by balancing the importance weights with optimized, data-dependent ``counterweights''.
This technique finds a natural application in sequential neural posterior estimation, which is an interesting direction for future work.
We also introduced a data-dependent rescaling of the coordinates, performed by a small neural network, that makes the posterior approximately standard normal while dealing with sharp parameter bounds and periodic boundary conditions.
This procedure is applicable in any simulation-based inference pipeline, to precondition the normalizing flow before expensive trainings are performed.
Finally, following Refs.~\cite{Lye2020, Mishra2021} we reduced overfitting by generating the training set on a low-discrepancy sequence.

We have made \texttt{labrador} open-source.\textsuperscript{\ref{ftn:github}} We tested our implementation on gravitational-wave inference with a quadrupole, aligned spin waveform model, on simulations and real data.
In terms of performance, \texttt{labrador} achieved a median importance-sampling efficiency of \medianeff, and excellent self-consistency in terms of probability--probability plots.
Regarding computational cost, training the model requires generating the training set ({\cpudays} CPU-core days, parallelizable, dominated by optimizing the reference waveform) and training the conditional normalizing flow (10 hours on an NVIDIA V100 GPU), plus other subdominant steps.
At inference time, \texttt{labrador} generates {\samplespersec} samples/s, which may then be reweighted for enhanced precision.

\begin{figure}
    \centering
    \includegraphics[width=\linewidth]{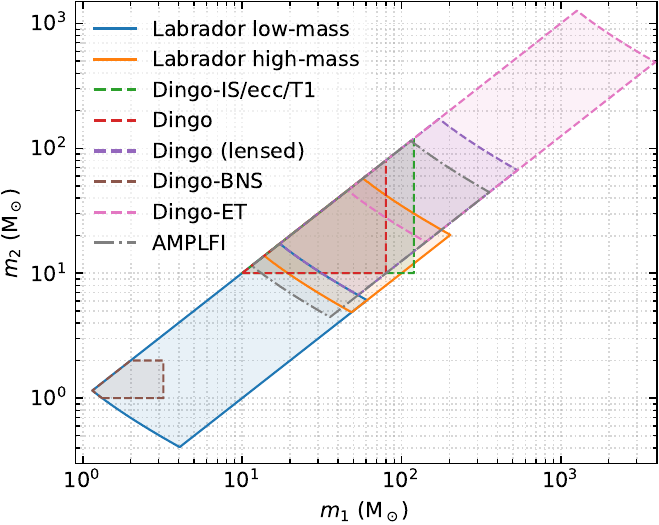}
    \caption{Mass ranges of various neural inference models. Thanks to the model's equivariance with respect to source parameter variations, \texttt{labrador} (this work) is able to fill a gap for sources with $\sim 3$--\SI{10}{M_\odot} components, whose duration rendered previous methods impractical.
    See Table~\ref{tab:models} for additional details.}
    \label{fig:mass_ranges}
\end{figure}

For comparison, various variants of \texttt{Dingo}, a state-of-the-art neural inference code, generally achieve higher median efficiencies (on the order of 10\%)
at the expense of using more complex models with a higher training cost (on the order of weeks on top-grade GPUs).
In Table~\ref{tab:models} we provide a comparison to neural posterior inference codes in current use, namely several variations of \texttt{Dingo} \cite{Dax2021, Dax2022, Dax2023, Dax2025, Gupte2025, Santoliquido2025, Caldarola2026, Spadaro2026}, and \texttt{AMPLFI} \cite{Chatterjee2024}.
We highlight some performance indicators including importance-sampling efficiency, training cost, and modeled physics.
\texttt{labrador} achieves a modest efficiency, albeit using significantly smaller models and training times than the other methods.
It should also be noted that many of those models account for more complex physical effects than this work---e.g.\ precession, higher harmonics, eccentricity, gravitational lensing, or variable detector networks---and also that the parameter ranges of the models differ.
Figure~\ref{fig:mass_ranges} shows the mass range of various models; notably, \texttt{labrador} extends the range of applicability of neural estimation towards long-duration, low-mass signals, which, thanks to our model's equivariances, are just as amenable to analysis as high-mass, short events.
It would be interesting to further investigate how the techniques we have introduced in this work influence the tradeoff between model complexity and performance.

\begin{table*}
    \centering
    \begin{ruledtabular}
        \begin{tabular}{lccccc}
            Model & Model size & Training cost (GPU) & Efficiency & Physics & Reference \\
            \hline
            \texttt{labrador} & \num{2.2e5} & \SI{10}{h} V100 & 1\% & --- & This work \cite{Roulet2026} \\
            \texttt{Dingo} & \num{1.3e8} & 16--\SI{18}{d} V100 & 1.4\% \cite{Kofler2025} & P & \cite{Dax2021, Dax2022} \\
            \texttt{Dingo-IS} & \num{1.3e8} & --- & 10\% & P, HM & \cite{Dax2023} \\
            \texttt{Dingo-BNS} & \num{1.3e8} & --- & 17\% & P, T & \cite{Dax2025} \\
            \texttt{Dingo-T1} & \num{1.6e8} & \SI{9.5}{d} 8 $\times$ A100 & 4.20\% & P, HM, N & \cite{Kofler2025} \\
            \texttt{Dingo} (eccentric) & --- & \SI{11}{d} A100 & 3.30\% & E, HM & \cite{Gupte2025} \\
            \texttt{Dingo} (Einsten Telescope) & --- & \SI{6}{d} A100 & 10.00\% & P, HM & \cite{Santoliquido2025} \\
            \texttt{Dingo} (lensed) & --- & \SI{14}{d} A100 & --- & P, HM, L & \cite{Caldarola2026} \\
            \texttt{Dingo-LISA} & \num{3.6e8} & \SI{10}{d} A100 & 22\% & HM & \cite{Spadaro2026} \\
            \texttt{AMPLFI} & \num{6e6} & \SI{20}{h} A40 & --- & --- & \cite{Chatterjee2024} \\
        \end{tabular}
    \end{ruledtabular}
    \caption{Comparison across neural posterior inference models.
    Model size is in terms of learnable parameters. The efficiency is the median importance sampling efficiency.
    Physical effects beyond quadrupolar, aligned spin waveforms are noted as P: precession; HM: higher modes; T: tides; N: variable detector networks; L: gravitational lensing, E: eccentricity.
    See Fig.~\ref{fig:mass_ranges} for the mass range of each model.
    }
    \label{tab:models}
\end{table*}

Our current implementation admits several extensions, namely accounting for a variable noise power spectrum and additional physical effects (spin-induced precession, tidal deformability, higher-order harmonics, eccentricity).
Precession and tidal deformability induce smooth perturbations to the signal amplitude and phase, and thus we expect our current heterodyning prescription to work without significant modifications. 
On the other hand, higher harmonics and eccentricity introduce richer frequency content and will require an improved treatment \cite{Leslie2021, Narola2024, Roulet2024}.
While we have reduced the computational cost of training, improving the accuracy achieved (as quantified by the importance-sampling efficiency) remains a challenge.
Perhaps the most promising avenue for increasing the efficiency in the short term will be to reduce the dimensionality of the posterior by splitting the space into intrinsic and extrinsic parameters and adapt efficient extrinsic-parameter marginalization techniques from the literature \cite{Veitch2015, Pankow2015, Islam2022, Roulet2024b, Mushkin2025} (within neural estimation, this approach has been successfully applied to the orbital phase parameter \cite{Dax2023, Santoliquido2025}).

The quality currently achieved by the amortized \texttt{labrador} posterior is likely sufficient for some uses, such as simulating large catalogs with realistic inference for forecasts \cite{Fishbach2018, Callister2020, Gray2020} or validation tests \cite{Farah2023}.
More demanding applications, such as electromagnetic follow-up, precision parameter inference, or population studies, would still require importance-sampling corrections.
Another use case of importance sampling the neural posterior is estimating the Bayesian evidence, which is the optimal detection statistic in a search.
As these methods mature and reweighting efficiencies improve, they may offer a practical route to incorporating advanced physical models into searches \cite{McIsaac2023, Wadekar2023, Wadekar2024, Schmidt2024b, Schmidt2024c, Dhurkunde2025, Wang2025, Nitz2026}. This prospect provides strong motivation for continued development of general and robust neural inference algorithms for gravitational-wave science.

%%%%%%%%%%%%%%%%%%%%%%%%%%%%%%%%%%%%%%%%
\section*{Acknowledgements}
%%%%%%%%%%%%%%%%%%%%%%%%%%%%%%%%%%%%%%%%
We thank Eliot Finch for designing the \texttt{labrador} logo (Fig.~\ref{fig:labrador}), and Carolina Cuesta L\'azaro, Stephen Green, Digvijay Wadekar and Mat\'ias Zaldarriaga for helpful discussions. 
JR thanks the Kavli Institute for Theoretical Physics for hospitality while part of this work was carried.
JR acknowledges support from the Sherman Fairchild Foundation and the Jonathan M.\ Nelson Center for Collaborative Research.
MC is funded by the European Union under the Horizon Europe's Marie Sk{\l}odowska-Curie project~101065440.
LMT is supported by NSF MPS-Gravity Awards 2207758 and 2513294, and by NSF Grant 2309200. 
KC was supported by NSF Grant PHY-2409001.
The authors are grateful for computational resources provided by the LIGO Laboratory and supported by National Science Foundation Grants PHY-0757058 and PHY-0823459. 
This material is based upon work supported by NSF's LIGO Laboratory which is a major facility fully funded by the National Science Foundation.

This research has made use of data or software obtained from the Gravitational Wave Open Science Center (\url{gwosc.org}), a service of LIGO Laboratory, the LIGO Scientific Collaboration, the Virgo Collaboration, and KAGRA.
LIGO Laboratory and Advanced LIGO are funded by the United States National Science Foundation (NSF) as well as the Science and Technology Facilities Council (STFC) of the United Kingdom, the Max-Planck-Society (MPS), and the State of Niedersachsen/Germany for support of the construction of Advanced LIGO and construction and operation of the GEO600 detector.
Additional support for Advanced LIGO was provided by the Australian Research Council.
Virgo is funded, through the European Gravitational Observatory (EGO), by the French Centre National de Recherche Scientifique (CNRS), the Italian Istituto Nazionale di Fisica Nucleare (INFN) and the Dutch Nikhef, with contributions by institutions from Belgium, Germany, Greece, Hungary, Ireland, Japan, Monaco, Poland, Portugal, Spain.
The construction and operation of KAGRA are funded by Ministry of Education, Culture, Sports, Science and Technology (MEXT), and Japan Society for the Promotion of Science (JSPS), National Research Foundation (NRF) and Ministry of Science and ICT (MSIT) in Korea, Academia Sinica (AS) and the Ministry of Science and Technology (MoST) in Taiwan.

\subsection*{Software used}

\noindent
\texttt{cogwheel} \cite{Roulet2019},
\texttt{gwpy} \cite{Macleod2021},
\texttt{HTCondor} \cite{Thain2005},
\texttt{nautilus} \cite{Lange2023},
\texttt{PyTorch} \cite{Paszke2019},
\texttt{sbi} \cite{Boelts2025},
\texttt{SciPy} \cite{Virtanen2020},
\texttt{XGBoost} \cite{Chen2016}.

\appendix

%%%%%%%%%%%%%%%%%%%%%%%%%%%%%%%%%%%%%%%%%%%
\section{Phenomenological waveform}
\label{app:waveform}
%%%%%%%%%%%%%%%%%%%%%%%%%%%%%%%%%%%%%%%%%%%

In this appendix we give details of the phenomenological model that we use in Eq.~\eqref{eq:calpha} to fit a reference waveform to the data, and provide a parametrization suitable for inputting it into the neural network.
As in the main text, a subindex $k$ labels detectors and $f$ labels frequencies.

The main component of the waveform model is the phase profile $\Phi_{kf}$. 
Per Eq.~\eqref{eq:calpha}, we seek to define it in terms of orthonormal coordinates, in which the mismatch metric is Euclidean.
We achieve this taking inspiration from the literature on geometric template banks \cite{Dhurandhar1994, Owen1996, Owen1999, Tanaka2000, Babak2006, Ajith2008, Brown2013, Roulet2019, Roy2019}.
We use
\begin{equation}\label{eq:phase_ansatz}
    \Phi_{kf}(\bm c)
        = \overline\Phi_{kf} + \sum_\alpha c_\alpha e_{kf\alpha}\,,
\end{equation}
where $\overline\Phi_{kf}$ is an average phase evolution (defined in Eq.~\eqref{eq:avg_phase} below), and $e_{kf\alpha}$ is a set of orthonormal bases:
\begin{align}
    \langle e_\alpha, e_\beta \rangle
    &\equiv \sum_{k f} W^2_{kf} \, e_{kf\alpha} \, e_{kf\beta}
    = \delta_{\alpha \beta}\,,\\
    W^2_{kf} &\propto \frac{\overline A^2_f}{S_{kf}} \Delta_f\,;
    \qquad\sum_{kf} W^2_{kf} = 1\,, \\
    \overline A_f &= f^{-7/6}\,.
\end{align}
where $S_{kf}$ is the noise PSD and $\Delta_f$ the frequency bin size.
Motivated by a Taylor expansion of the phase near the peak of the likelihood, the inner product is weighted by the expected squared signal-to-noise ratio in each frequency bin and detector.
Orthonormality implies that in the $\bm c$ coordinates the Hessian of the log-likelihood is proportional to the identity, and so in the high-SNR limit the likelihood resembles an isotropic Gaussian, with width $\sigma_{\bm c} = 1 / \text{SNR}$ \cite{Roulet2019}.

We obtain the orthonormal basis through a combination of QR and singular-value decompositions of post-Newtonian phase profiles. From the 1.5 post-Newtonian approximation we have\footnote{\label{note:HL}For concreteness we illustrate with a Hanford--Livingston network; generalization to other detectors is straightforward.}
\begin{equation}
    \Phi_{kf} = \sum_n P_{kfn} \, a_n\,,
\end{equation}
where
\begin{center}
    \begin{tabular}{ccl}
        $n$ & $P_{kfn}$ & $a_n$ \\
        \hline
        0 & $\delta_{k\mathrm{H}}$           & $\varphi_{\mathrm{H}}$  \\
        1 & $\delta_{k\mathrm{L}}$           & $\varphi_{\mathrm{L}}$  \\
        2 & $- 2\pi f\,\delta_{k\mathrm{H}}$ & $t_{\mathrm{H}}$     \\
        3 & $-2\pi f\,\delta_{k\mathrm{L}}$  & $t_{\mathrm{L}}$     \\
        4 & $f^{-5/3}$                       & $a_{\mathrm{0PN}}(\mathcal M)$   \\
        5 & $f^{-1}$                         & $a_{\mathrm{1PN}}(\mathcal M, \eta)$   \\
        6 & $f^{-2/3}$                       & $a_{\mathrm{1.5PN}}(\mathcal M, \eta, \chi_{1z}, \chi_{2z})$ \\
    \end{tabular}
\end{center}
\vspace{10pt}
Above, $\{P_{kfn}\}$ span the space of phase profiles and have known coefficients $a_n$, but they are not orthogonal.
We now find an orthonormal basis.

%-----------------------------------------
\subsection{Extrinsic parameters}

We first do a QR decomposition of the (weighted) $P$ matrix, to obtain orthogonalized phase and time coordinates:
\begin{equation}
    W_{kf}P_{kfn} = \sum_m Q_{kfm}R_{mn}\,,
\end{equation}
where $Q$ is orthogonal (thinking of $kf$ as a joint index) and $R$ is upper triangular:
\begin{align}
    \sum_{kf} Q_{kfn}Q_{kfm} &= \delta_{n m}\,, \\
    R_{m>n} &= 0\,.
\end{align}
We define $e_{kfn} \equiv Q_{kfn} / W_{kf}$ for the first functions describing pure detector phase and time shifts,  $n < 2N_{\rm det}$.
For example, $Q_{kf0}$ generates a phase shift at Hanford, $Q_{kf1}$ a phase shift at Livingston,  $Q_{kf2}$ a time-and-phase shift at Hanford that is orthogonal to a pure phase shift, etc.
We use a QR decomposition here because it keeps the first bases pure phase shifts, which is important as this allows us to maximize the phase coefficients analytically when optimizing the likelihood.

%--------------------------------------
\subsection{Intrinsic parameters}

The remaining components encode the dependence on intrinsic parameters.
As it turns out, for inspiral-dominated (low-mass) systems the 1PN and 1.5PN terms are comparable in size but quite degenerate.\footnote{We focus this discussion on low-mass systems, for which heterodyning is most important in terms of removing variability from the signals.
Regardless, we apply the same procedure uniformly in parameter space.}
In other words, we need to include the 0PN, 1PN, 1.5PN to accurately modeling the signal, but the space of phase profiles associated with the intrinsic parameters is effectively 2-dimensional rather than 3-dimensional  \cite{Cutler1994}.
This can be seen e.g.\ in that the low-mass template banks of \cite[Table I]{Roulet2019} are 2-dimensional.

To have fewer optimization parameters and limit the ability of the phenomenological model to produce unphysical waveforms, we choose the remaining bases (that describe the effect of intrinsic parameters) as the principal components of a set of examples of phase profiles orthogonalized to phase and time shifts.
We generate many examples (labeled by $i$) of physical parameters, compute the associated PN coefficients $a_{n i}$ and subtract the average
\begin{equation}
    \overline a_n = \frac 1N \sum_i^N a_{n i}\,.
\end{equation}
We build a matrix of examples of weighted phase residuals orthogonalized to time and phase shifts, as
\begin{equation}
\begin{split}
    M_{kfi}
    &= W_{kf} (\Phi^\perp_{kfi} - \overline \Phi_{kf}^\perp) \\
    &= \!\sum_{n, m \geq 2 N_\mathrm{det}}\! Q_{kfm} \, R_{mn} \, (a_{n i} - \overline a_n)\,,
\end{split}
\end{equation}
where $n,m$ only run over the components orthogonal to the phase and time bases (strictly this makes sense for $m$; $n$ follows because $R$ is upper triangular). We then do an SVD of this:
\begin{equation}
    M_{kfi} = \sum_{n} U_{kf\alpha} \, D_{\alpha} \, V_{i\alpha}\,.
\end{equation}
Above, $U$ is orthogonal and defines the remaining bases ($e_{kf\alpha} = U_{kf\alpha} / W_{kf}$ for $\alpha \geq 2 N_\mathrm{det}$), of which we keep only the first two (these contain most of the information, as mentioned earlier).
The average phase in Eq.~\eqref{eq:phase_ansatz} is
\begin{equation}\label{eq:avg_phase}
    \overline \Phi_{kf}
    = \! \sum_{n \geq 2 N_\mathrm{det}} \! P_{kfn} \, \overline a_{n}\,.
\end{equation}

%----------------------------------------
\subsection{Feature design}
\label{sec:network-friendly-parametrization}

The reference-waveform parameters, which form part of the data representation, serve as context for the conditional normalizing flow.
To make them amenable for the neural network, we express certain features in relative rather than absolute form.

Changes in  the distance, orbital phase, or merger time produce a global rescaling, phase shift, or time shift across all detectors.
The information relevant for sky localization is instead contained in the amplitude ratios, phase differences, and time delays between detectors.
For example, we replace $(t_{\mathrm{H}}, t_{\mathrm{L}})$ with $t_{\mathrm{H}} - t_{\mathrm{L}}$, which defines a ring of sky locations.
We express phase lags as $\cos(\phi_{\mathrm{H}} - \phi_{\mathrm{L}})$ and $\sin(\phi_{\mathrm{H}} - \phi_{\mathrm{L}})$ to avoid $2\pi$ discontinuities.
Similarly, we encode the network signal-to-noise ratio as $A = \sqrt{A_{\mathrm{H}}^2 + A_{\mathrm{L}}^2}$ and use amplitude ratios $A_k / A$ to preserve directional information.\textsuperscript{\ref{note:HL}}
Removing the overall time and phase from the data representation enforces an exact equivariance under time and phase translations.

%%%%%%%%%%%%%%%%%%%%%%%%%%%%%%%%%%%%%%%
\section{Architecture details}
\label{app:architecture}
%%%%%%%%%%%%%%%%%%%%%%%%%%%%%%%%%%%%%%

In this appendix we describe the architecture of the models used in this paper.
These choices are easily customizable in \texttt{labrador}.

For both the low- and high-mass models in Table~\ref{tab:simulation_prior} we use a training set of \num{e7} simulations placed on a Halton sequence.

For the $\mu(d), \Sigma^{-1/2}(d)$ models in the rescaler (\S\ref{sec:rescaling}), we have used a multilayer perceptron with 4 hidden layers of 32 neurons, totaling 8078 trainable parameters, with a SiLU activation function.
We have used the AdamW optimizer \cite{Loshchilov2019} with a learning rate scheduler.

For the posterior estimator, we have used a rational-quadratic neural spline flow \cite{Durkan2019} consisting of 5 transforms with 10 bins (spline nodes) each, with 64 hidden features, totaling \num{208448} trainable parameters.
We have used the implementation in \texttt{sbi}~\cite{Boelts2025}, modified to use low-discrepancy batches and weighted simulations per Eq.~\eqref{eq:prior_weights}.

\bibliography{main}

\end{document}